\documentclass[twocolumn,showpacs,preprintnumbers,amsmath,nofootinbib,
               floatfix,amssymb]{revtex4}

\usepackage{graphicx}
\usepackage{bm}
\usepackage{dcolumn}
\begin{document}

\title{Numerical investigation of photon creation in a
       three-dimensional resonantly vibrating cavity: TE-modes}

\author{Marcus Ruser\footnote[1]{Electronic address: Marcus.Ruser@physics.unige.ch}}
\affiliation{D{\'e}partement de Physique Th{\'e}orique, Universit{\'e}
de Gen{\`e}ve, 24 quai E. Ansermet, CH-1211 Gen{\`e}ve 4 Switzerland}

\begin{abstract}
The creation of TE-mode photons in a three-dimensional perfectly conducting 
cavity with one resonantly vibrating wall is studied numerically. 
We show that the creation of TE-mode photons in a rectangular cavity
is related to the production of massive scalar particles on a
time-dependent interval. The equations of motion are solved
numerically which allows to take into account the intermode coupling. 
We compare the numerical results with analytical predictions and
discuss the effects of the intermode coupling in detail.  
The numerical simulations reveal that photon creation in a
three-dimensional resonantly vibrating cavity can be maximized by
arranging the size of the cavity such that certain conditions are
realized. In particular, the creation of TE-mode photons in the 
lowest frequency mode $(1,1,1)$ is most efficient in a non-cubic cavity 
where the size of the non-dynamical dimensions is roughly 11 times
larger than the size of the dynamical dimension. We discuss this
effect and its relation to the intermode coupling in detail.
\end{abstract}

\pacs{03.65.-w, 03.70.+k, 12.20.Ds, 42.50.Lc}

\maketitle
        
\section{Introduction}
In 1948 Casimir \cite{Casimir:1948} predicted an attractive force between two 
perfectly conducting plates (ideal mirrors). This so-called Casimir
effect \cite{Plunien:1986, Bordag:2001,Mostepanenko:1997,Milton:2004} 
caused by the change of the zero point energy of the 
quantized electromagnetic field in the presence of boundaries 
has been verified experimentally with high accuracy 
\cite{Lamoreaux:1997,Mohideen:1998,Roy:1999a,Roy:1999b,Bressi:2002}. The 
existence of the Casimir force \cite{Lamoreaux:2005} acting on
macroscopic boundaries confirms the reality of quantum vacuum 
fluctuations and their potential influence even on macroscopic scales. 

Besides the change of the zero point energy of the quantum vacuum provoked by 
static boundary conditions a second and even more fascinating feature of the 
quantum vacuum appears when considering dynamical, i.e time-dependent boundary 
conditions. The quantum vacuum responds to time-varying boundaries with the 
creation of real particles (photons) out of virtual quantum vacuum fluctuations. 
This effect, usually referred to as dynamical or non-stationary Casimir effect 
\cite{Dodonov:2001a}, has gained growing interest during recent years.

 A scenario of particular interest are so-called vibrating cavities 
\cite{Lambrecht:1996} where the distance between two parallel mirrors 
changes periodically in time. The possibility of resonance effects
between the mechanical motion of the mirror and the quantum vacuum
leading to an even exponential growth of the particle 
occupation numbers for the resonance modes makes this configuration the most 
promising candidate for an experimental verification of the dynamical 
Casimir effect.

For a one-dimensional vibrating cavity this effect has been studied in 
numerous works \cite{Dodonov:1993,Dodonov:1996,Dodonov:1996a,Ji:1997,
Schuetzhold:1998, Dodonov:1998, Klimov:1997,Fu:1997, Chizhov:1997,Law:1994,Cole:1995,
Meplan:1996,Dalvit:1998,Rog:2001,Andreata:2000,Llave:1999,Dalvit:1999,Ji:1998} showing
that the total energy inside a resonantly vibrating cavity increases
exponentially in time. 

The more realistic case of a three-dimensional cavity is studied in 
\cite{Crocce:2001,Crocce:2002, Dodonov:1995,Dodonov:2003,
Mundarain:1998,Dodonov:2001,Dodonov:1998a,Dodonov:1998b}. 
The important difference between one- and higher-dimensional cavities is that the 
frequency spectrum in only one spatial dimension is equidistant while it is in general 
non-equidistant for more spatial dimensions. An equidistant spectrum 
yields strong intermode coupling whereas in case of a non-equidistant
spectrum only a few or even no modes may be coupled allowing for 
exponential photon creation in a resonantly vibrating
three-dimensional cavity \cite{Dodonov:1996,Crocce:2001,Crocce:2002}. 
Without intermode coupling the equations of motion for the field modes 
reduce to harmonic oscillators with 
time-dependent frequency. Particle creation can then be investigated 
by using an approach based on Schr{\"o}dinger scattering theory \cite{Sassaroli:1994}. 
Even though for higher-dimensional cavities the problem can be reduced to a single 
harmonic oscillator in some special cases \cite{Dodonov:1996} the intermode coupling 
cannot be neglected in general \cite{Crocce:2001,Dodonov:2001}. 
(See also the discussion of the work 
\cite{Sassaroli:1994} in section IX of \cite{Dodonov:1996}.) 

Field quantization inside cavities with non-perfect boundary conditions has been 
studied in, e.g., \cite{Schaller:2002a,Schaller:2002b} and corrections due to 
finite temperature effects are treated in 
\cite{Plunien:2000,Jing:2000,Schuetzhold:2002}. 
The interaction between the quantum vacuum and the (classical) dynamics
of the cavity has been investigated in 
\cite{Law:1994, Law:1995,Golestanian:1997,Cole:2001}
and an approach to the dynamical Casimir effect based on stationary
walls but time-dependent conductivity properties is discussed in
\cite{Crocce:2004}.

The electromagnetic field inside a dynamical cavity can be decomposed 
into components corresponding to the electric field parallel or 
perpendicular to the moving mirror. It is then possible to
introduce vector potentials for each polarization, 
transverse electric (TE) and transverse magnetic (TM) 
\cite{Maia Neto:1994,Maia Neto:1996,Mundarain:1998}. 
The equations of motion for TE-modes in a dynamical rectangular cavity 
are equivalent to the equations of motion for a scalar field 
with (time-dependent) Dirichlet boundary conditions 
\cite{Crocce:2001,Crocce:2002}. More complicated boundary 
conditions, so-called generalized Neumann boundary conditions, emerge when studying 
TM-modes \cite{Maia Neto:1994, Mundarain:1998}. In most of the works cited above only 
TE-polarizations are treated. For recent work dealing also with 
TM-polarizations see \cite{Crocce:2002,Crocce:2005}.

The aim of the present work is to study photon creation in a
vibrating three-dimensional cavity fully numerically taking the intermode 
coupling into account. We show that the equations of motion describing
the dynamics of the transverse electric modes (TE) in a 
dynamical rectangular cavity correspond to the equations of motion for
a massive scalar field on a time-dependent interval (one-dimensional cavity).  
Thereby the wave number of the TE-modes associated with the
non-dynamical cavity dimensions is identified with the mass of the
scalar field. Creation of TE-polarized photons can then be studied
with a formalism presented and tested for a massless scalar field 
in a one-dimensional cavity in \cite{Ruser:2005,Ruser:2006a}. 
Even though the method of \cite{Ruser:2005,Ruser:2006a} is valid for a 
variety of boundary conditions it is not directly applicable to 
generalized Neumann boundary conditions which involve 
a time-derivative appearing when studying TM-modes. 
For other recent numerical work see also 
\cite{Antunes:2003,Li:2002,Fedotov:2005}.

The paper is organized as follows. In section II we present the
equations of motion for TE-modes in a three-dimensional rectangular
cavity and show that they correspond to the equations of motion for 
a massive scalar field in a one-dimensional cavity. 
The formalism for studying the dynamical Casimir effect for a 
massive scalar field on a time-dependent interval 
numerically is reviewed in section III. Some analytical results 
obtained for TE-mode photons are summarized in section IV. 
We present and interpret the numerical results in section V and
discuss their consequences for 
photon creation in three-dimensional vibrating cavities in section VI. 
We conclude in section VII and discuss some details about the numerics
in the appendix.

\section{Equations of motion for TE-modes in a rectangular dynamical cavity}
The dynamics of the transverse electric modes (TE-modes) inside a 
rectangular ideal (i.e. perfectly conducting) cavity of dimensions 
$\{(0,l_x),(0,l_y)(0,l_z)\}$ is described by the 
wave (Klein-Gordon) equation \footnote{We are using units with $\hbar=c=1$.}
\begin{equation}
[\partial_t^2-\triangle]\Phi(t,{\bf x})=0
\label{KG equation}
\end{equation}
with the massless scalar field $\Phi(t,{\bf x})$ subject to
Dirichlet boundary conditions at all walls of the cavity
\cite{Crocce:2001,Crocce:2002}.

The $x$-dimension of the cavity is assumed to be dynamical 
with the right wall following a prescribed trajectory 
$l(t) \equiv l_x(t)$. At any moment in time the field can be expanded as
\begin{equation}
\Phi(t,{\bf x})=\sum_{\bf n} q_{\bf n}(t)\phi_{\bf n}(t,{\bf x})
\label{field expansion}
\end{equation}
with canonical variables $q_{\bf n}(t)$ and functions 
\begin{eqnarray}
\phi_{\bf n}(t,{\bf x}) &=&\sqrt{\frac{2}{l(t)}}\sin\left[\frac{n_x\pi}{l(t)}x\right]
\sqrt{\frac{2}{l_y}}\sin\left[\frac{n_y\pi}{l_y}y\right]\nonumber \\
&~&\times\sqrt{\frac{2}{l_z}}\sin\left[\frac{n_z\pi}{l_z}z\right]
\end{eqnarray} 
ensuring Dirichlet boundary conditions at the positions of the cavity walls
\cite{Crocce:2001}. The functions $\phi_{\bf n}(t,{\bf x})$ form an
orthonormal and complete set of instantaneous eigenfunctions 
of the Laplacian $\triangle$ with time-dependent eigenvalues
\begin{equation}
\Omega_{{\bf n}}(t)=\pi\sqrt{\left(\frac{n_x}{l(t)}\right)^2+
\left(\frac{n_y}{l_y}\right)^2+\left(\frac{n_z}{l_z}\right)^2}.
\end{equation}
Each field mode is labeled by three integers $n_x,n_y,n_z=1,2,...$ 
for which we use the abbreviation ${\bf n}=(n_x,n_y,n_z)$. 

Inserting the expansion \eqref{field expansion} into the 
field equation \eqref{KG equation}, multiplying it with 
$\phi_{\bf m}(t,x)$ and integrating over the spatial dimensions 
leads to the equation of motion for the canonical variables 
$q_{\bf n}(t)$ \cite{Crocce:2001}:
\begin{eqnarray}
\ddot{q}_{\bf n}(t) &+& \Omega^2_{\bf n}(t)q_{\bf n}(t) 
+2\sum_{\bf m}M_{\bf mn}(t)\dot{q}_{\bf m}(t) 
\label{deq}
\\
&+&\sum_{\bf m}\left[\dot{M}_{\bf mn}(t) - N_{\bf nm}(t)\right]
q_{\bf m}(t)=0.
\nonumber
\end{eqnarray}
The time-dependent coupling matrices $M_{\bf nm}(t)$ and $N_{\bf nm}(t)$ 
are given by \cite{Crocce:2001}
\begin{eqnarray}
M_{\bf nm}&=&\int_0^{l(t)}dx\,\dot{\phi}_{\bf n}\phi_{\bf m}
\label{coupling matrix}\\
&=&\frac{\dot{l}(t)}{l(t)}
\left\{\begin{array}{ll}
(-1)^{n_x+m_x}\frac{2n_xm_x}{m_x^2-n_x^2}\delta_{n_ym_y}\delta_{n_zm_z}
& {\rm if} \;n_x\neq m_x \nonumber \\
0 & {\rm if} \; n_x=m_x
\end{array}\right .
\end{eqnarray}
and
\begin{equation}
N_{\bf nm}=\sum_{\bf k}M_{\bf nk}M_{\bf mk}.
\label{N matrix}
\end{equation}
During the dynamics of the mirror the time evolution of a
field mode ${\bf n}$ may be coupled to (even infinite many) other
modes ${\bf m}$ via the time-dependent coupling matrix $M_{\bf nm}(t)$.
In Eq.~(\ref{deq}) for a given mode $(n_x,n_y,n_z)$ the coupling 
matrix (\ref{coupling matrix}) yields couplings
of $q_{(n_x,n_y,n_z)}$ to $q_{(m_x,n_y,n_z)}$ and 
$\dot{q}_{(m_x,n_y,n_z)}$, i.e. only summations over $m_x$ appear. 
Modes with different quantum numbers in the $y-$ and $z-$ directions 
are not coupled and the quantum numbers corresponding to the
non-dynamical dimensions enter the equations of motion only globally.
Therefore we can identify $q_n(t) \equiv q_{(n_x,n_y,n_z)}(t)$ and  
\begin{equation}
\Omega_n(t) \equiv \Omega_{(n_x,n_y,n_z)}(t)  =
\sqrt{\left[\frac{n\pi}{l(t)}\right]^2 + k_\|^2}
\label{freq translation}
\end{equation}
with $n\equiv n_x$ and the wave number 
\begin{equation}
k_\|=\pi\sqrt{\left(\frac{n_y}{l_y}\right)^2 + \left(\frac{n_z}{l_z}\right)^2}   
\label{def of mass}
\end{equation}
associated with the non-dynamical cavity dimensions.  

Because all summations over ${\bf m}=(m_x,m_y,m_z)$ involving the
coupling matrix (\ref{coupling matrix}) reduce to summations over a 
single quantum number $m$, Eq. (\ref{deq}) is equivalent to the 
differential equation describing a real massive scalar field on 
a time-dependent interval $[0,l(t)]$ (one-dimensional cavity) 
when $k_{\|}$ is identified with the mass of the field 
\cite{Ruser:2006a}. 
As pointed out in \cite{Crocce:2002} the number of created TE-mode 
photons equals the number of created Dirichlet scalar
particles in a three-dimensional cavity. Consequently, the number of 
TE-mode photons created in a three-dimensional cavity equals the 
number of scalar particles of ``mass'' $k_{\|}$ created in a 
one-dimensional cavity $[0,l(t)]$. Photon production in TE-modes 
in a three-dimensional cavity can therefore be studied numerically with the 
formalism presented in \cite{Ruser:2005,Ruser:2006a}. 

\section{The formalism}
Quantization is achieved by replacing the set of classical canonical
variables $\{q_{n},p_{m}\}$ with the corresponding operators
$\{\hat{q}_{n},\hat{p}_{m}\}$ and demanding the usual 
equal-time commutation relations. Furthermore, the Heisenberg 
picture is adopted from now on. The relation between the canonical 
variable $q_{n}$ and the canonical momentum is given by 
$p_{n}=\dot{q}_{n}+\sum_{m}q_{m}M_{mn}$. Assuming that the cavity is at rest 
for times $t\le 0$ the coupling matrix vanishes and equation 
\eqref{deq} reduces to the equation of a harmonic oscillator 
with constant frequency $\Omega_{n}^0\equiv \Omega_{n}(t \le 0)$.  
Consequently 
\begin{equation}
\hat{q}_{n}(t\le 0)=\frac{1}{\sqrt{2\Omega^0_{n}}}
\left[\hat{a}_{n}e^{-i\Omega^0_{n}t} +
      \hat{a}^\dagger_{n}e^{i\Omega^0_{n}t}\right] 
\end{equation}
with frequency
\begin{equation}
\Omega_n^0=\frac{1}{l_0}\sqrt{\left(n\pi\right)^2+M^2}
\end{equation}
where $l_0=l(0)$ and we have introduced the dimensionless 
``mass parameter'' $M=l_0 \,k_{||}$. The time-independent annihilation 
and creation operators $\hat{a}_{n},\hat{a}_{n}^\dagger$ associated with the 
particle notion for $t\le 0$ are subject to the 
commutation relations
\begin{equation}
\left[\hat{a}_{n},\hat{a}_{m}\right]=
\left[\hat{a}_{n}^\dagger,\hat{a}_{m}^\dagger\right]=0\,,\;
\left[\hat{a}_{n},\hat{a}_{m}^\dagger\right]=\delta_{nm}.
\end{equation}
The initial vacuum state $|0,t\le 0\rangle$ is defined by 
\begin{equation}
\hat{a}_n |0,t\le 0\rangle=0\;\forall \;n.
\end{equation} 
When the cavity dynamics is switched on at $t=0$ and the wall
follows the prescribed trajectory $l(t)$ field modes are coupled 
due to the non-vanishing coupling matrix $M_{nm}$. 
To account for the coupling, the operator $\hat{q}_{n}$ may be expanded as 
\cite{Ruser:2006a}  
\begin{equation}
\hat{q}_{n}(t\ge 0)=\sum_m\frac{1}{\sqrt{2\Omega_{m}^0}}
\left[\hat{a}_{m} \epsilon_{n}^{({m})}(t) + 
      \hat{a}_{m}^\dagger\epsilon_{n}^{({m})^*}(t)\right]
\label{expansion of q}
\end{equation}
with complex functions $\epsilon_{n}^{({m})}(t)$ satisfying 
Eq.~\eqref{deq}. 
If the motion ceases and the wall is at rest again for $t \ge t_1$ the operator 
$\hat{q}_{n}(t\ge t_1)$ takes the form \footnote{Here $l(t_1)=l_1$
  is assumed to be arbitrary. For an oscillating cavity, however, it is natural
  to consider times $t_1$ after which the dynamical wall has returned
  to its initial position.} 
\begin{equation}
\hat{q}_{n}(t\ge t_1)=\frac{1}{\sqrt{2\Omega^1_{n}}}
\left[\hat{A}_{n}e^{-i\Omega^1_{n}(t-t_1)} +
      \hat{A}^\dagger_{n}e^{i\Omega^1_{n}(t-t_1)}\right] 
\label{final q}
\end{equation}
with $\Omega_n^1\equiv \Omega_n(t\ge t_1)$ and 
annihilation and creation operators 
$\hat{A}_{n}, \hat{A}_{n}^\dagger$ 
corresponding to the particle notion for $t\ge t_1$. 
The final vacuum state $|0,t\ge t_1\rangle$ is defined by
\begin{equation}
\hat{A}_n |0,t\ge t_1\rangle=0\;\forall \;n.
\end{equation} 
The initial state particle operators $\hat{a}_{n},
\hat{a}_{n}^\dagger$ are linked to the final state particle operators 
$\hat{A}_{n}, \hat{A}_{n}^\dagger$ by the Bogoliubov transformation 
\begin{equation}
\hat{A}_{n}=\sum_{m}\left[{\cal A}_{mn}(t_1)\hat{a}_{m}
+{\cal B}_{mn}^*(t_1)\hat{a}_{m}^\dagger\right]
\end{equation}
and the number of particles (photons) created in a mode $n$ during 
the motion of the wall is given by the expectation value of the number operator 
$\hat{A}^\dagger_{n}\hat{A}_{n}$ associated with the particle
notion for $t \ge t_1$ with respect to the initial
vacuum state $|0,t\le 0\rangle$:
\begin{eqnarray}
N_{n}(t_1)&=&\langle 0,t\le 0|\hat{A}_{n}^\dagger\hat{A}_{n}|0,t\le 0\rangle
\nonumber \\
&=&
\sum_{m}|{\cal B}_{mn}(t_1)|^2.
\label{particle number}
\end{eqnarray}
The total number of created particles as the sum of $N_n(t_1)$ 
over all quantum numbers $n$
\begin{equation}
N(t_1)=\sum_nN_n(t_1)=\sum_n\sum_m|{\cal B}_{mn}(t_1)|^2
\end{equation}
is in general ill-defined and requires appropriate regularization.
This can be done most easily by introducing an explicit frequency
cut-off which also simulates non-ideal boundary conditions for high frequency
modes \cite{Schuetzhold:1998}. As a matter of course, 
such a frequency cut-off has to be used in the numerical simulations. 

In order to calculate ${\cal B}_{mn}(t_1)$ we introduce 
auxiliary functions $\xi_n^{(m)}(t)$ and $\eta_n^{(m)}(t)$ via
\cite{Ruser:2006a}
\begin{equation}
\xi_n^{(m)}(t)= \epsilon_n^{(m)}(t)
+\frac{i}{\Omega_n^0}\left[\dot{\epsilon}_n^{(m)}(t)+
\sum_kM_{kn}(t)\epsilon_k^{(m)}(t)\right],
\label{def xi}
\end{equation}
\begin{equation}
\eta_n^{(m)}(t)= \epsilon_n^{(m)}(t)
-\frac{i}{\Omega_n^0}\left[
\dot{\epsilon}_n^{(m)}(t)+\sum_kM_{kn}(t_1)\epsilon_k^{(m)}(t)\right]
\label{def eta}.
\end{equation}
Using the second order differential equation (\ref{deq}) for 
$\epsilon_n^{(m)}(t)$ it is easily shown that those functions 
satisfy the following system of coupled 
first-order differential equations \cite{Ruser:2006a}:
\begin{eqnarray}
\dot{\xi}_{n}^{({m})}(t)&=&-i\left[a^+_{nn}(t)\xi_{n}^{({m})}(t)-
a^-_{nn}(t)\eta_{n}^{({m})}(t)\right]\nonumber\\
&-&\sum_{k}\left[c^-_{nk}(t)\xi_{k}^{({m})}(t)+
c^+_{nk}(t)\eta_{k}^{({m})}(t)\right],
\label{deq for xi}
\end{eqnarray}
\begin{eqnarray}
\dot{\eta}_{n}^{({m})}(t)&=&-i\left[a^-_{nn}(t)\xi_{n}^{({m})}(t)-
a^+_{nn}(t)\eta_{n}^{({m})}(t)\right]\nonumber\\
&-&\sum_{k}\left[c^+_{nk}(t)\xi_{k}^{({m})}(t)+
c^-_{nk}(t)\eta_{k}^{({m})}(t)\right]
\label{deq for eta}
\end{eqnarray}
with 
\begin{eqnarray}
a_{nn}^\pm(t)&=&\frac{\Omega_{n}^0}{2}
\left\{1 \pm \left[\frac{\Omega_{n}(t)}{\Omega_{n}^0}\right]^2\right\},
\label{def a matrix} \\
c_{kn}^\pm(t)&=&\frac{1}{2}\left[M_{{nk}}(t) \pm 
\frac{\Omega_{n}^0}{\Omega_{k}^0}M_{kn}(t)\right].
\label{def c matrix}
\end{eqnarray}
For the coupling matrix (\ref{coupling matrix}) one finds in particular
\begin{eqnarray}
c_{kn}^\pm(t)=-\frac{\dot{l}(t)}{l(t)}(-1)^{k+n}\frac{kn}{n^2-k^2}
\left[1\mp \frac{\Omega_n^0}{\Omega_k^0}\right]
\end{eqnarray}
if $n\neq k$ and $c_{nn}^\pm(t)=0$.
The advantage of this system of first-order differential
equations relies on the fact that, besides the time-dependent
frequency $\Omega_{n}(t)$, only the coupling matrix $M_{nk}$
enters but neither its square $N_{nk}$ nor its time derivative
$\dot{M}_{nk}$. 

By matching Eq.~(\ref{expansion of q}) with Eq.~(\ref{final q}) 
for $\hat{q}_n(t)$ and the corresponding expressions 
for $\hat{p}_n(t)$ at $t=t_1$ one finds the relations 
\cite{Ruser:2006a}
\begin{equation}
{\cal A}_{mn}(t_1)=\frac{1}{2}\sqrt{\frac{\Omega_n^1}{\Omega_n^0}}
\left[\Delta_n^+(t_1)\xi_n^{(m)}(t_1) +
  \Delta_n^-(t_1)\eta_n^{(m)}(t_1)\right],
\label{Bogoliubov A}
\end{equation}
  \begin{equation}
{\cal B}_{mn}(t_1)=\frac{1}{2}\sqrt{\frac{\Omega_n^1}{\Omega_n^0}}
\left[\Delta_n^-(t_1)\xi_n^{(m)}(t_1) +
  \Delta_n^+(t_1)\eta_n^{(m)}(t_1)\right]
\label{Bogoliubov B}
\end{equation}
with
\begin{equation}
\Delta_n^+(t)=\frac{1}{2}\left[ 1 \pm \frac{\Omega_n^0}{\Omega_n(t)}\right].
\end{equation}

Demanding that the field be in its vacuum state \\
$|0,t\le 0\rangle$ 
as long as the mirror is at rest implies 
${\cal A}_{mn}(0)=\delta_{mn}$
and ${\cal B}_{mn}(0)=0$. Accordingly the initial conditions for
$\xi_n^{(m)}(t)$ and $\eta_n^{(m)}(t)$ read 
\begin{equation}
\xi_{n}^{({m})}(0)=2\delta_{mn}\,,\;\eta_{n}^{({m})}(0)=0.
\end{equation}
By means of Eq.~(\ref{Bogoliubov B}) the number of created massive 
scalar particles, or equivalently the
number of created TE-mode photons, at time $t=t_1$ can now be calculated by solving
the system of differential equations formed by Eqs. (\ref{deq for xi}) and
(\ref{deq for eta}) numerically using standard numerics. 
For this we truncate the infinite sums by introducing a cut-off
quantum number $k_{\rm max}$ to make the system of differential
equations suitable for numerical treatment. The system is evolved up
to a final time $t_{\rm max}$ and the particle number 
(\ref{particle number}) is calculated for several times in between,
i.e. we interprete $t_1$ as a continuous variable such that the
particle number (\ref{particle number}) becomes a continuous function
of time 
\footnote{Potential problems inherent in this procedure like the
  appearance of discontinuities in the velocity of the mirror motion
  occurring when calculating the particle number for times $t$ for
  which $\dot{l}(t)\neq 0$ are discussed in \cite{Ruser:2006a} in
  detail. We come back to this in section V. 
}. 
Consequently, the stability of the numerical results has to be
guaranteed which means that for the lowest modes $n$ the numerical 
values for $N_n(t)$ remain practically unchanged under variation 
of $k_{\rm max}$. 
More details regarding the numerics are collected in the appendix.    

\section{Known analytical results}
In what follows, we consider the periodic trajectory
\begin{equation}
l(t)=l_0\left[1+\epsilon\sin(\omega\,t)\right]\;,\;\epsilon
\ll 1,
\label{sine motion}
\end{equation}
for which it was found in \cite{Crocce:2001} that two modes 
$l$ and $k$ are coupled whenever one of the conditions given by
\begin{equation}
\omega=|\Omega_l^0 \pm \Omega_k^0|
\label{coupling condition}
\end{equation}
is satisfied
\footnote{Here and in the following we have translated the results for
  three-dimensional cavities to the case of massive scalar particles 
  according to Eq.~(\ref{freq translation}). 
}. In a resonantly vibrating cavity  
$\omega=2\Omega_n^0$ with not one of those conditions  
fulfilled the number of TE-mode photons 
created in the resonant mode $n$ increases exponentially in time
\cite{Crocce:2001}: 
\begin{equation}
N_n(t)=\sinh^2(n\,\gamma_n \, \epsilon \,t)\;\;{\rm with}\;\;
\gamma_n=\frac{n}{2\,\Omega_n^0}\left(\frac{\pi}{l_0}\right)^2.
\label{massive particle number}
\end{equation}
By means of multiple scale analysis the authors of \cite{Crocce:2001} 
also studied the resonance case $\omega=2\Omega_n^0$ 
with two coupled modes $n$ and $k$ satisfying  
\begin{equation}
3\Omega_n^0=\Omega_k^0.
\label{nk coupling}
\end{equation}
For the particular case $n=1$ and $k=5$ analytical expressions 
for the number of TE-mode photons are derived in \cite{Crocce:2001}. 
Given a mode $n$ we can couple it to a particular mode $k$ by tuning 
the mass $M$ (or equivalently $k_\|$) such that the condition 
(\ref{nk coupling}) is fulfilled. 
It is important to note that coupling between modes does occur even if  
Eq.~(\ref{coupling condition}) is detuned, i.e. if 
Eq.~(\ref{coupling condition}) is satisfied by the frequencies $\Omega_k^0$ and
$\Omega_l^0$ only approximately. The particular case of two modes $n$ and $k$
satisfying 
\begin{equation}
(3+\kappa)\Omega_n^0=\Omega_k^0
\label{detuned coupling condition}
\end{equation}
without additional couplings to higher modes was studied in \cite{Dodonov:2001}.
For sufficiently small $\kappa$ (i.e., $\kappa < \epsilon$) the two modes 
$n$ and $k$ are still resonantly coupled and the number of particles produced in 
both modes increasing exponentially with time.  

\section{Numerical results}
\subsection{Preliminary remarks}
In \cite{Ruser:2005,Ruser:2006a} we have employed the same formalism 
to study the creation of massless scalar particles in a 
one-dimensional vibrating cavity numerically. 
In this case the numerical results agree with analytical
predictions obtained under the assumption $\epsilon \ll 1$ 
demonstrating the reliability of the numerics. The extension to
massive scalar fields is straightforward. 
We set $l_0=1$, i.e. all physical quantities with dimensions are 
measured with respect to the length scale $l_0$ and dimensionless
quantities are used throughout. 
The amplitude of the oscillations is fixed to $\epsilon=0.001$ 
guaranteeing accordance between the numerical results and the
analytical predictions in the massless case
\cite{Ruser:2005,Ruser:2006a}.   

As mentioned before, we calculate the particle number for
arbitrary times even though the analytical expressions we are
comparing the numerical results with are valid only for times after
which the dynamical wall has returned to its initial position. In
\cite{Ruser:2005} it is shown that (for a vibrating cavity) 
this leads to oscillations in the particle number which are of
negligibly small amplitude when the amplitude of the cavity oscillations itself
is small ($\epsilon \ll 1$). We will briefly come back to this 
question later on. Furthermore, when calculating the particle number at
times $t$ for which $\dot{l}(t)\neq 0$ the used particle definition requires 
a matching of the solutions to expressions corresponding to the static 
cavity $\dot{l}(t)=0$. This discontinuity in the velocity of the
mirror trajectory may give rise to spurious contributions to the total
particle number. The cut-off $k_{\rm max}$ automatically ensures that
the total particle number remains finite because it automatically
smoothes the motion. For a more detailed discussion see
\cite{Ruser:2006a} where it is shown that the influence of such 
discontinuities is negligibly small in the case
of a cavity vibrating with sufficiently small amplitudes $(\epsilon=0.001)$.
(For a detailed discussion of how the initial discontinuity
in the mirror motion (\ref{sine motion}) affects the particle creation
see also \cite{Ruser:2005}.) The numerical results which are presented
and discussed in the following are practically not affected by the 
above mentioned effects.   

\subsection{Main resonance $\omega=2\Omega_1^0$}
\begin{figure}
\includegraphics[height=5.6cm]{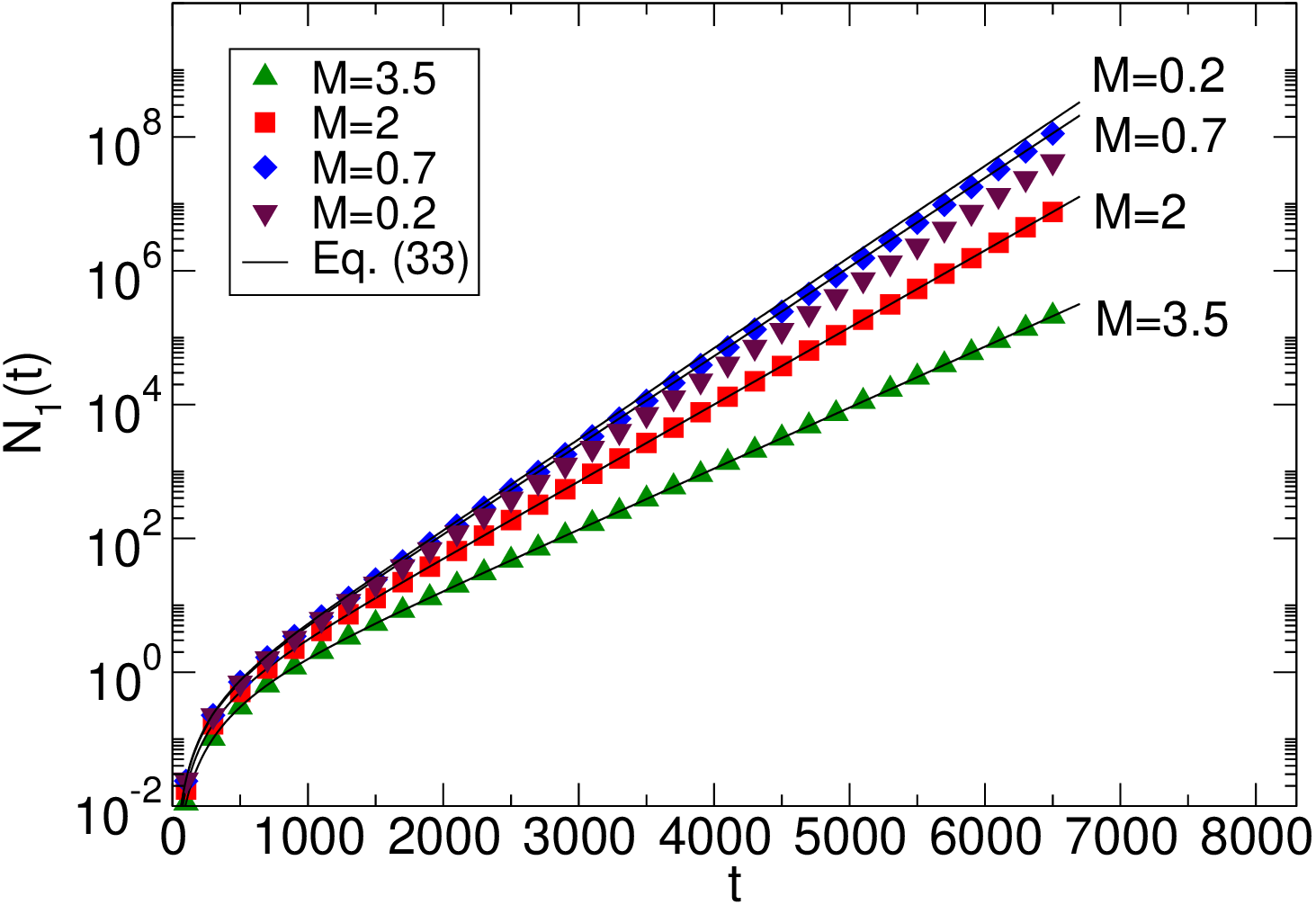}
\caption{(color online) Number of particles created in the resonance
  mode $n=1$ for mass parameters $M=0.2, 0.7$, $2$, and $3.5$  
  in comparison with the analytical prediction (\ref{massive particle
  number}).   
\label{figure 1}}
\end{figure}

\begin{figure}
\includegraphics[height=5.6cm]{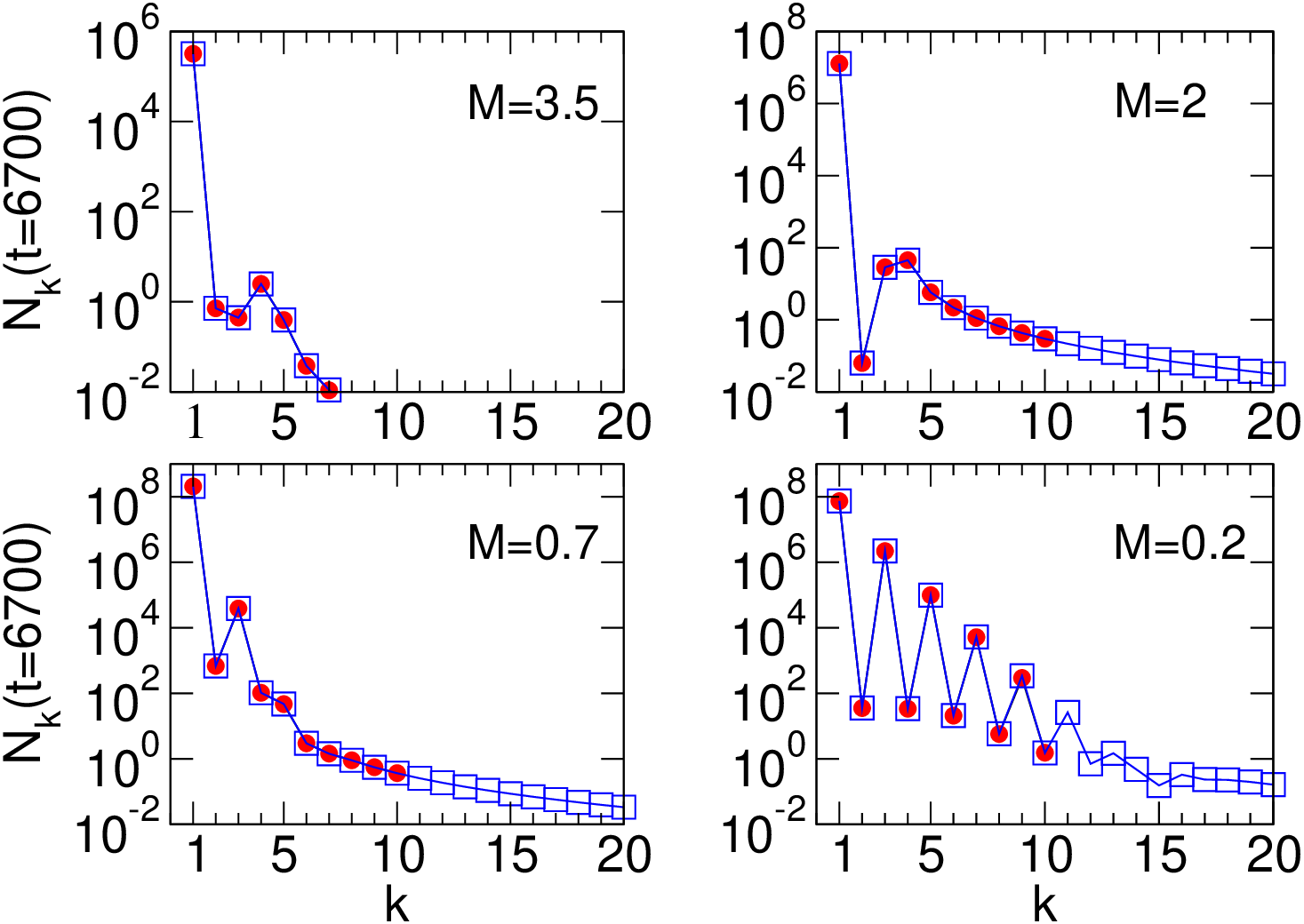}
\caption{(color online) Particle spectrum for different mass parameters $M=3.5, 2,
0.7$ and $M=0.2$ at time $t=6700$ corresponding to 
Fig.~\ref{figure 1}. The spectra are shown for $k_{\rm max}=10$ (dots) 
and $k_{\rm max}=20$ (squares) to demonstrate numerical stability.  
\label{figure 2}}
\end{figure}

\begin{figure}
\includegraphics[height=5.6cm]{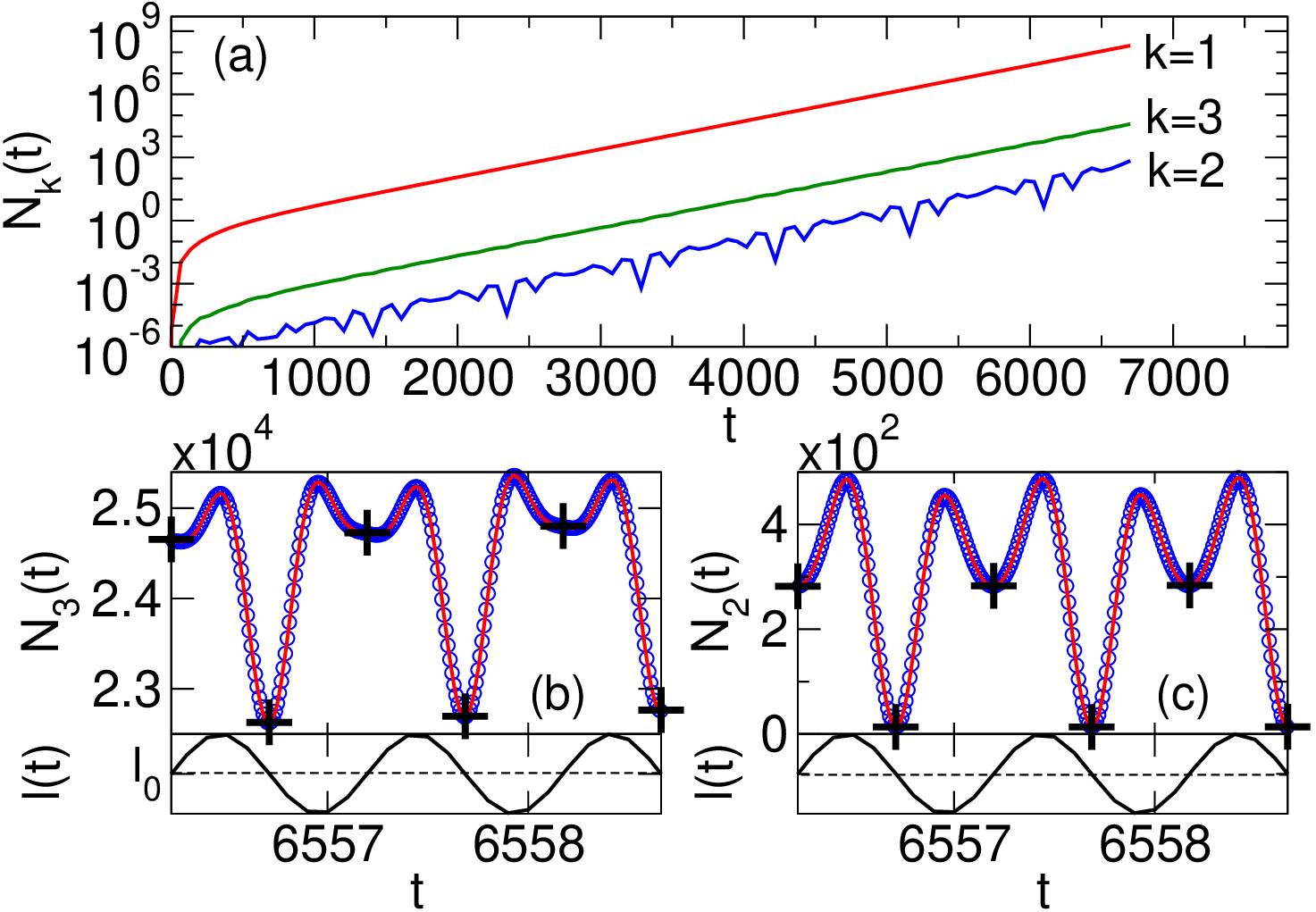}
\caption{(color online) (a) Number of particles created in the modes $k=1$, $2$ and
  $3$ for the mass parameter $M=0.7$ corresponding to the spectra shown in
  Fig. \ref{figure 2}. 
  Part (b) shows $N_3(t)$ and part (c) $N_2(t)$ in each case for the two resolutions 
  $\Delta t=0.01$ (circles) and $\Delta t=0.005$ (solid lines). The
  particle numbers calculated for times at which the mirror has returned
  to its initial position are accented by ``${\bf +}$'' and the  
  background motion is shown for comparison as well.   
\label{figure 3}}
\end{figure}

In Fig.~\ref{figure 1} the number $N_1(t)$ of particles 
created in the resonant mode $k=1$ is shown for masses  
$M=0.2,0.7,2$ and $3.5$ and compared to the analytical prediction 
Eq.~(\ref{massive particle number}) 
\footnote{In this section we use the general notion ``particles'' for 
massive scalar particles, or equivalently, TE-mode
photons. Furthermore we call $M$ the mass of the particle, having
in mind that it corresponds to the wave number $k_{\|}$ 
for TE-mode photons.}. For $M=0.7,2$ and $3.5$ the numerical results
are well described by Eq.~(\ref{massive particle number}) which is
valid provided that the resonant mode $n=1$ is not coupled to other 
modes. In case of the mass $M=0.2$ the numerical result for $N_1(t)$ 
disagrees with the analytical prediction 
(\ref{massive particle number}). This will be discussed 
in the following in detail. Figure~\ref{figure 2} shows the corresponding 
particle spectra at time $t=6700$. 
One infers that for $M=3.5, 2$ and $0.7$ the 
mode which becomes excited most is indeed the resonant mode $n=1$. However, also 
higher modes become excited but the corresponding particle numbers 
are several orders of magnitude smaller than the number of particles 
created in the resonant mode. For $M=0.7$, for example, the mode $k=3$
is clearly excited. Figure~\ref{figure 3} shows the number of 
particles created in the modes $k=1,2$ and $3$ for the 
mass parameter $M=0.7$ in detail. 
The difference in the numerical values of $N_1$ and $N_3$ is so large
that the contribution of $N_3$ to the total particle number is negligible such that 
$N\simeq N_1$. From Fig.~\ref{figure 3}~(a) one could 
conclude that $N_2$, i.e. the number of particles created in the
mode $k=2$, behaves in the same way as $N_3$ but shows superimposed 
oscillations. However, Figs.~\ref{figure 3}~(b) and (c) provide a 
more detailed view on the time evolution of $N_k(t)$ for modes $k=2$ and $3$. In 
Fig.~\ref{figure 3} (a) the resolution in which the numerical results are shown 
is not sufficient in order to resolve the details which are visible in 
Figs.~\ref{figure 3} (b) and (c). These high resolution pictures
reveal that $N_3$  increases exponentially in time with 
oscillations superimposed on an average particle number whereas 
$N_2$ itself oscillates strongly with an amplitude negligibly 
small compared to $N_1$. 

The small scale oscillations in the particle numbers are correlated
with the periodic motion of the mirror which we have depicted in 
Figs. \ref{figure 3} (b) and (c) as well. But one infers that the particle
numbers show oscillations also when the expectation value
(\ref{particle number}) is calculated only for times at which
the mirror has returned to its initial position $l_0$. Therefore the 
oscillations cannot be traced back exclusively to the fact that  
Eq.~(\ref{particle number}) is calculated for arbitrary
times which may be considered as unphysical.  

\begin{figure}
\includegraphics[height=5.6cm]{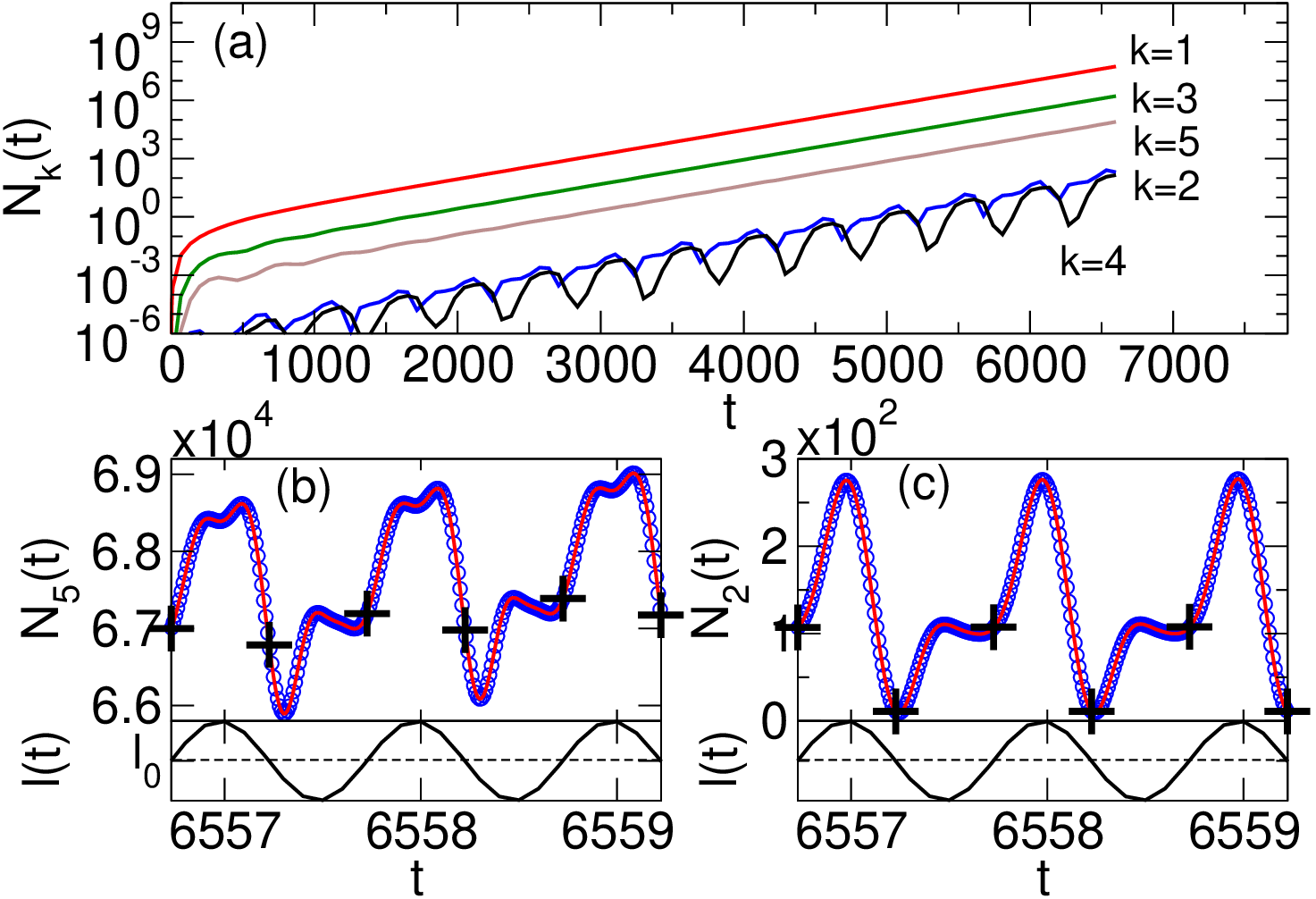}
\caption{(color online) (a) Number of particles created in the modes 
$k=1,2,3,4$ and $5$ for the 
mass parameter $M=0.2$ corresponding to the spectrum shown in Fig.~\ref{figure 2}.   
Part (b) shows $N_5(t)$ and part (c) $N_2(t)$ in each case for the two resolutions 
$\Delta t=0.01$ (circles) and $\Delta t=0.005$ (solid lines).
\label{figure 4}}
\end{figure}

The observation that also higher modes become excited (even though they are very much
suppressed) is explained by the fact that two modes $k$ and $l$ are coupled 
even if Eq.~(\ref{coupling condition}) is not exactly satisfied by the two 
frequencies $\Omega_k^0$ and $\Omega_l^0$. For $M=0.7$ the equation 
$3\Omega_1^0=\Omega_k^0$ has no solution for 
integer $k$. Thus taking Eq.~(\ref{coupling condition}) as an exact equation only the 
resonant mode should become excited and particle creation should take place in the 
mode $n=1$ exclusively. Inserting $M=0.7$ one finds the solution $k\sim3.07$ 
which is apparently close enough to the integer value $k=3$ to excite that mode. 
For smaller values of $M$ the solution of $3\Omega_1^0=\Omega_k^0$ 
approaches the value $k=3$ and one has to expect that for sufficiently 
small values of $M$ the mode coupling 
becomes again so strong that Eq.~(\ref{massive particle number}) does no longer 
describe the numerical results. This is the case for $M=0.2$ yielding 
$k\sim 3.005$ for which a strong coupling between the modes 
$n=1$ and $k=3$ occurs. Furthermore, from 
Eq.~(\ref{coupling condition}) and the coupling 
of $n=1$ and $k=3$ follows $2\Omega_1^0+\Omega_3^0=\Omega_l^0$ which has 
$l\sim 5.004$ as solution, i.e. the mode $k=3$ is coupled to the mode
$l=5$. In the same way the mode $5$ is coupled to the mode $7$.
Thus interpreting Eq.~(\ref{coupling condition}) 
as $\omega\simeq |\Omega_k^0 \pm \Omega_l^0|$ explains the
numerically computed particle spectrum (cf. Fig. \ref{figure 2})
which shows similar features as the spectrum obtained for the massless 
case (cf Fig.~4 (b) of \cite{Ruser:2005}). One observes that also even 
modes become excited (like also for $M=0.7$) which is not the case for 
$M=0$ \cite{Ruser:2005}. These modes are dragged by the strongly
excited modes (odd modes) and rapidly oscillate (like $N_2$ for
$M=0.7$) with an amplitude several orders of magnitude smaller
compared to $N_1$, $N_3$ and $N_5$. In Fig.~\ref{figure 4} we 
show the number of particles created in the modes $k=1$ to $5$ for $M=0.2$ to 
illustrate the just-stated. 
As for $M=0.7$ the number of particles created in the odd modes increases 
exponentially showing oscillations superimposed on an average particle number while 
the number of particles created in the even modes $k=2$ and $k=4$ consists of 
oscillations only with amplitudes much smaller compared to the number
of particles created in the odd modes. As in Fig.~\ref{figure 3} the
strongly oscillating behavior of the particle numbers for 
even modes is visible in high time resolution only [part (c) of Fig.~\ref{figure 4}].

The fact that mode coupling occurs even if Eq.~(\ref{coupling condition}) 
is not satisfied exactly is well known. We can rewrite the expression 
$3\Omega_n\simeq \Omega_k^0$ to get $(3+\kappa)\Omega_n^0=\Omega_k^0$ 
[Eq.~(\ref{detuned coupling condition})]. 
As mentioned at the end of the former section it was shown for this case 
in \cite{Dodonov:2001} that for sufficiently small $\kappa$ the modes 
$n$ and $k$ are still resonantly coupled, provided that no coupling to higher 
modes exists. 
However, the case of two detuned coupled modes does not apply to
the scenario discussed here. Decreasing the detuning, i.e. 
reducing the value of $M$, does not only strengthen the coupling between the modes 
$n=1$ and $k=3$ which would lead to an exponential growth of the particle number in 
both modes but also enhances the coupling strength to higher modes $k=5,7,...$
because the frequency spectrum becomes equidistant as $M\rightarrow 0$
(cf, e.g., \cite{Ruser:2005,Dodonov:1996}). The convergence of the
numerical results towards the analytical expressions for the massless
case is demonstrated below.  

\begin{figure}
\includegraphics[height=5.6cm]{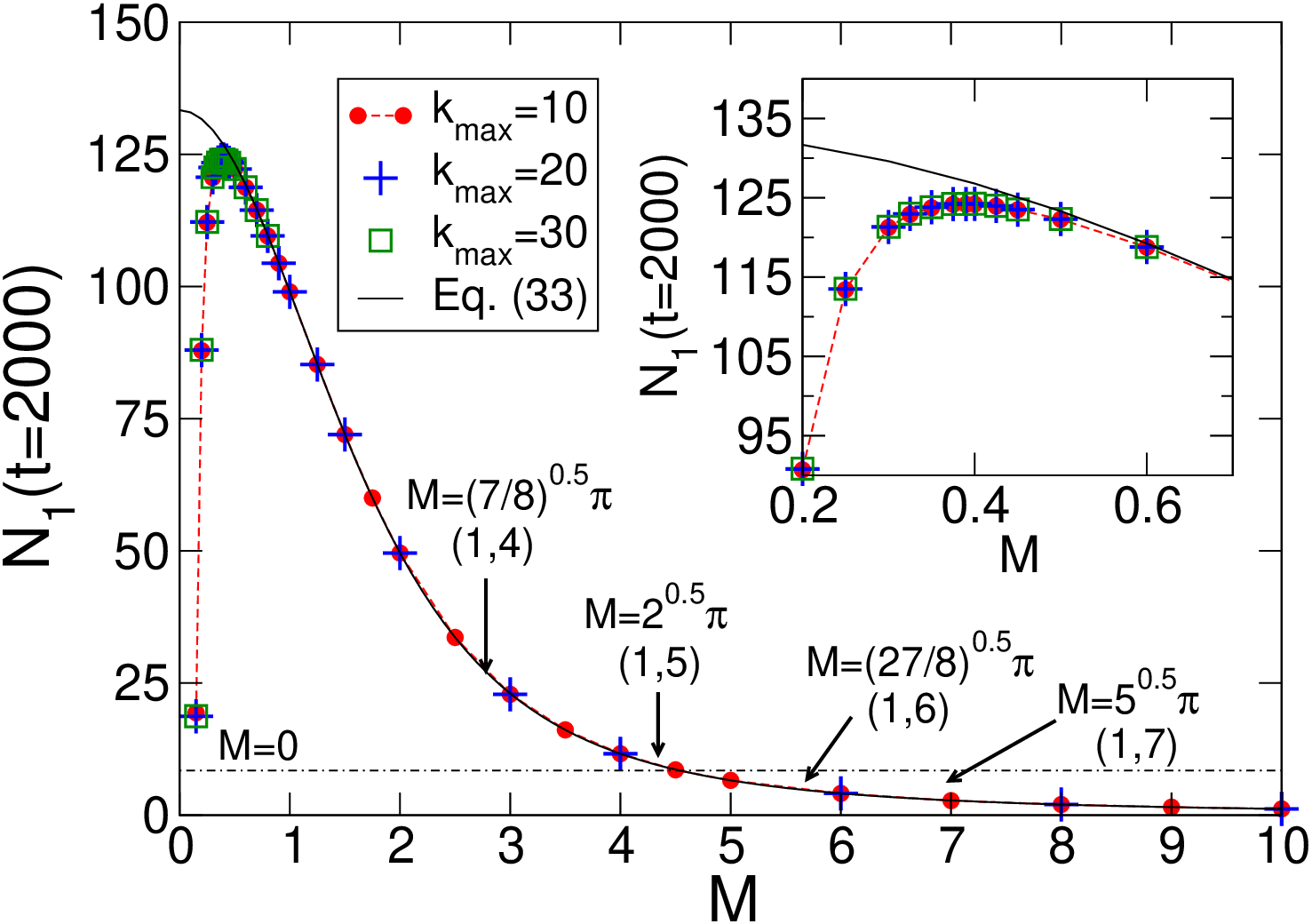}
\caption{(color online) Number of particles created in the resonance
  mode $n=1$ at time $t=2000$ as a function of the 
mass parameter $M$. The solid line shows the analytical prediction 
Eq.~(\ref{massive particle number}). Arrows pointing towards particular mass values 
of $M$ mark masses for which Eq.~(\ref{massive particle number}) is not valid 
because of exact intermode coupling. The coupled modes are given in brackets 
$[(1,k)]$. No numerical results are shown in the plot for those cases.
Most of the numerical results are shown for different values of the cut-off 
$k_{\rm max}$ to underline stability.  
\label{figure 5}}
\end{figure}

To study in more detail how the number of produced particles depends on the 
mass we performed numerical simulations for a wide range of values 
for $M$. The results are summarized in Fig.~\ref{figure 5} in a 
"mass spectrum" where the number of particles created in the resonant
mode $N_1(t=2000)$ is plotted as a function of $M$ and compared 
to the analytical prediction  Eq.~(\ref{massive particle number}). 
Particular values of $M$ for which Eq.~(\ref{coupling condition}) 
gives integer solutions, i.e. exact (un-detuned) intermode coupling, are
marked by arrows and the values of $M$ are indicated. Numerical 
results for these values are not included in the spectrum. Cases with 
exact coupling will be discussed later on.

\begin{figure}
\includegraphics[height=5.6cm]{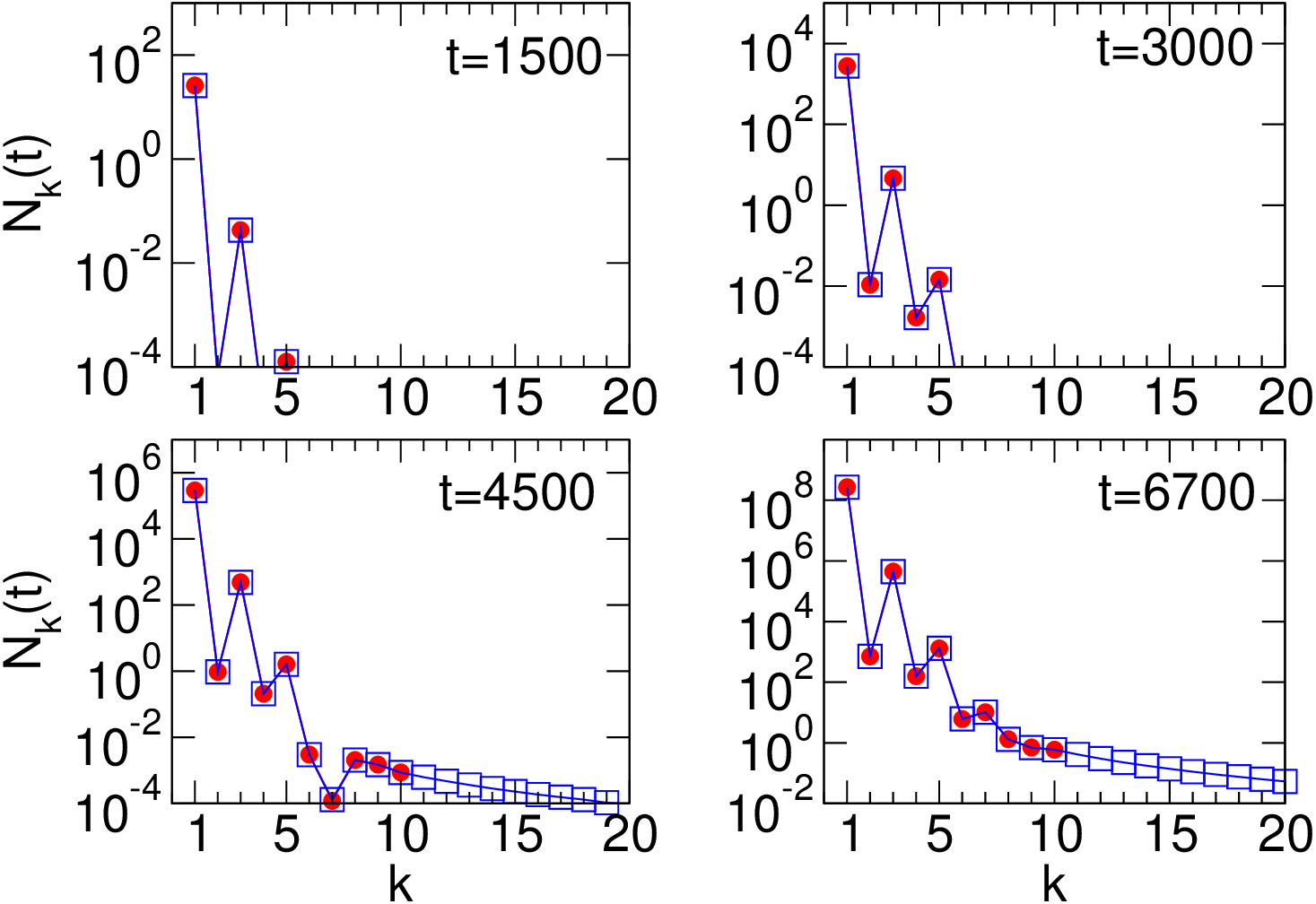}
\caption{(color online) Particle spectra for mass parameter $M=0.4$ 
at times $t=1500, 3000, 4500$ and
$6700$. Each spectrum is shown for values $k_{\rm max}=10$ (dots) and $k_{\rm max}=20$
(squares) to indicate numerical stability. 
\label{figure 6}}
\end{figure}

\begin{figure}
\includegraphics[height=5.3cm]{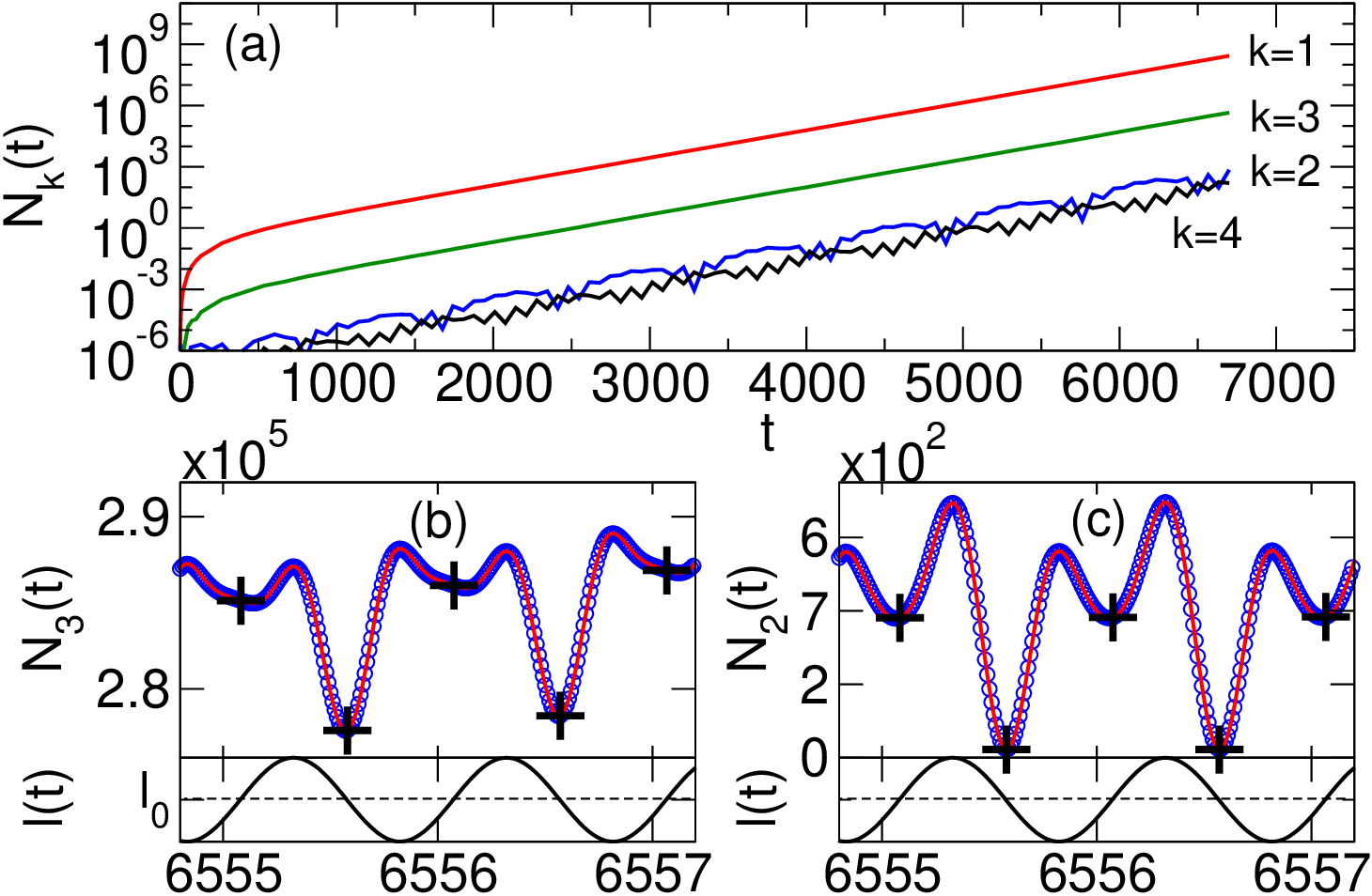}
\caption{(color online) (a) Number of particles created in the 
modes $k=1,2,3$ and $4$ for the mass 
parameter $M=0.4$ corresponding to the spectra shown in Fig.~\ref{figure 6}.    
Part (b) shows $N_3(t)$ and part (c) $N_2(t)$ in each case for the two resolutions 
$\Delta t=0.01$ (circles) and $\Delta t=0.005$ (solid lines).
\label{figure 7}}
\end{figure}

The numerical values for $N_1$ perfectly agree with the analytical prediction 
(\ref{massive particle number}) for values of $M$ larger than roughly 
$M=0.6$. For masses smaller than this threshold value the number of created 
particles is smaller compared to the analytical prediction. 
The mass spectrum exhibits a maximum at around $M\sim 0.4$, i.e. 
particle production in the resonant mode is most efficient for this 
particular mass. When $M < 0.4$ the number of created particles drops down and 
approaches the $M=0$ result. The appearance of a maximum in the mass
spectrum is clear from the above discussion. For the particular value 
$M=0.4$ the equation $3\Omega_1^0=\Omega_k^0$ 
leads to a value $k=3.02$ which is close enough to the integer solution 
$k=3$ to couple this mode strongly to the resonant mode but on the other hand 
coupling to higher modes is still suppressed. Figure~\ref{figure 6}
shows the particle spectrum obtained for $M=0.4$ for different times 
and in Fig.~\ref{figure 7} the time evolution of the number of
particles created in the modes $k=1,2,3$ and $4$ is plotted. 
For the even modes $k=2$ and $4$ the same oscillating behavior is 
observed as for $M=0.7$ and $M=0.2$. 
The coupling of the mode $k=3$ to the mode $n=1$ results in a damping 
of the resonant mode and consequently the number of particles
produced in the mode $n=1$ is smaller than the value predicted by 
Eq. (\ref{massive particle number}). 

\begin{figure}
\includegraphics[height=5.6cm]{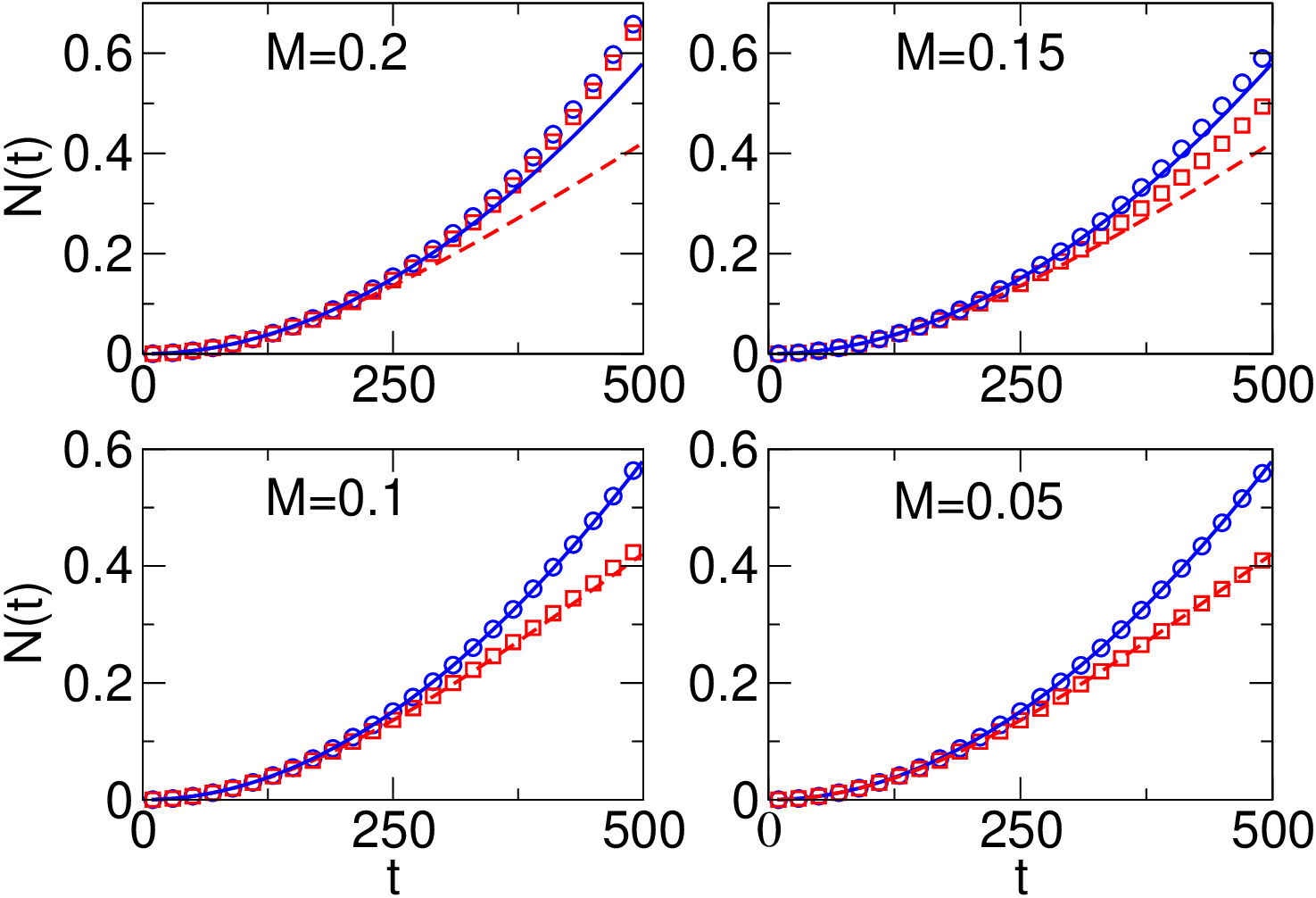}
\caption{(color online) Total particle number $N$ (circles) and number 
of particles created in the 
mode $k=1$ $N_1$ (squares) for mass parameters $M=0.2, 0.15, 0.1$ and $0.05$ together 
with the analytical predictions for the massless case Eq.~(6.5) (dashed line) and 
Eq.~(6.10) (solid line) of \cite{Dodonov:1996} to demonstrate the convergence of 
the solutions towards the $M=0$ case (see also Fig.~4 of \cite{Ruser:2005}). 
The cut-off parameter $k_{\rm max}=30$ was used in the simulations. 
\label{figure 8}}
\end{figure}

For increasing masses larger than $M=0.4$ the excitation of higher modes 
becomes more and more suppressed (cf Fig.~\ref{figure 2}). 
Accordingly the numerical results match the analytical expression 
\eqref{massive particle number} predicting that the number of 
created particles decreases with increasing mass. 
Decreasing the mass below $M=0.4$ enhances the strength of the
intermode coupling which results in a damping of the resonant mode $n=1$.
Consequently the number of particles produced in the mode $n=1$ 
(and also the total particle number) is smaller than predicted
analytically. When studying the 
limit $M\rightarrow 0$ the numerical results should converge towards
the well known results for the massless case where all odd modes are 
coupled \cite{Ruser:2005,Dodonov:1996}. 
This is demonstrated in Fig.~\ref{figure 8} where the total particle
number and the number of particles created in the resonant mode $n=1$ 
are depicted for $M=0.2,0.15, 0.1$ and $0.05$ up to $t=500$
and compared with the analytical predictions for $M=0$ 
\cite{Dodonov:1996} (see also Figs.~4~(a) and (b) of \cite{Ruser:2005}). 
While for $M=0.2$ the total particle number $N(t)$ is still mainly given by 
$N_1(t)$ a divergency between $N(t)$ and $N_1(t)$ starts to become
visible for $M=0.15$, i.e. the influence of the intermode coupling
gains importance. For $M=0.1$ the numerical 
results are close to the analytical $M=0$-results and are practically 
identical to them for $M=0.05$. 

\begin{figure}
\includegraphics[height=5.6cm]{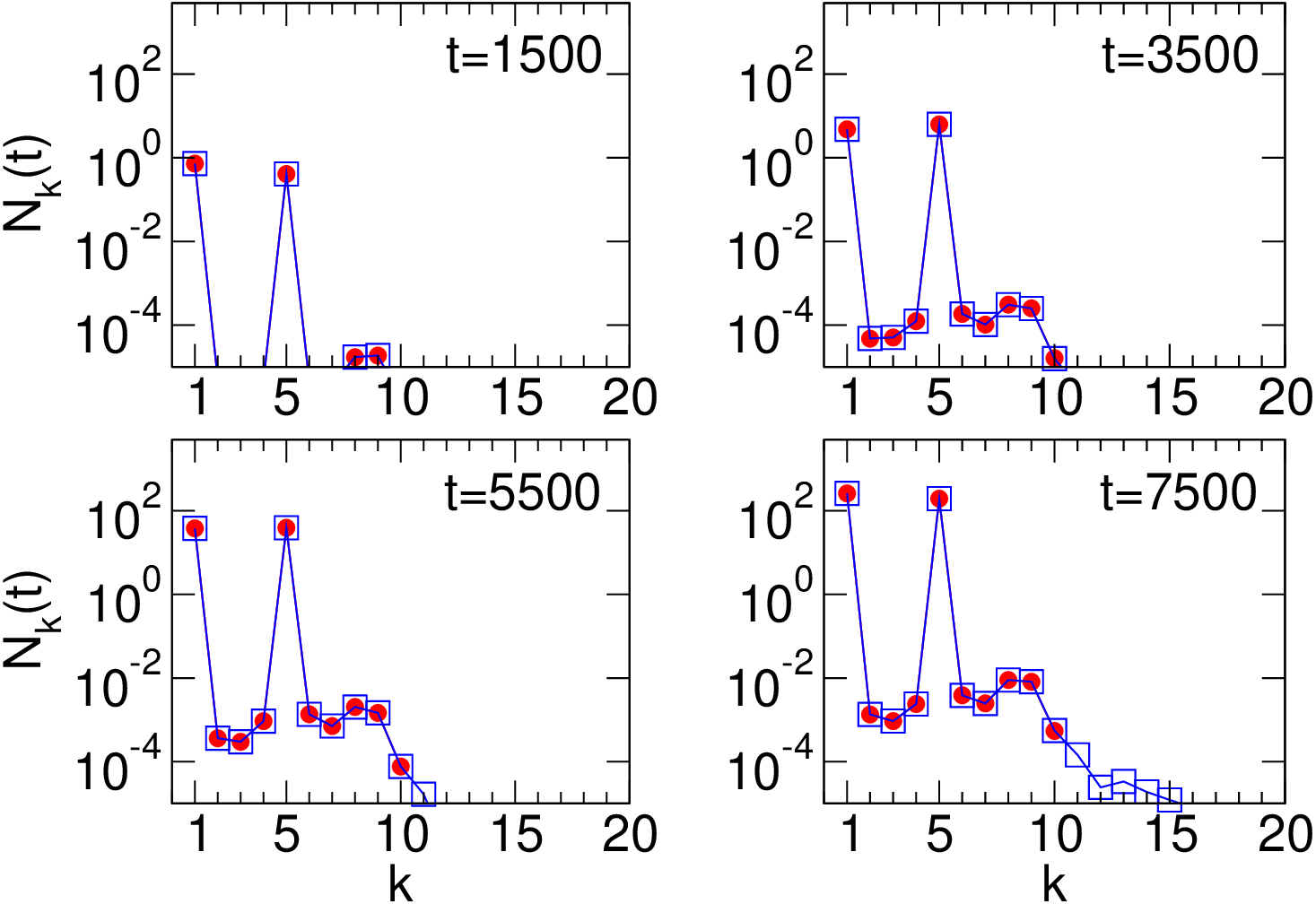}
\caption{(color online) Particle spectra for 
$\omega=2\Omega_1^0$ and mass parameter $M=\sqrt{2}\pi$ 
yielding exact coupling between the modes $n=1$ and
$k=5$. Dots correspond to $k_{\rm max}=10$ and squares to $k_{\rm max}=20$.
\label{figure 9}}
\end{figure}

\begin{figure}
\includegraphics[height=5.6cm]{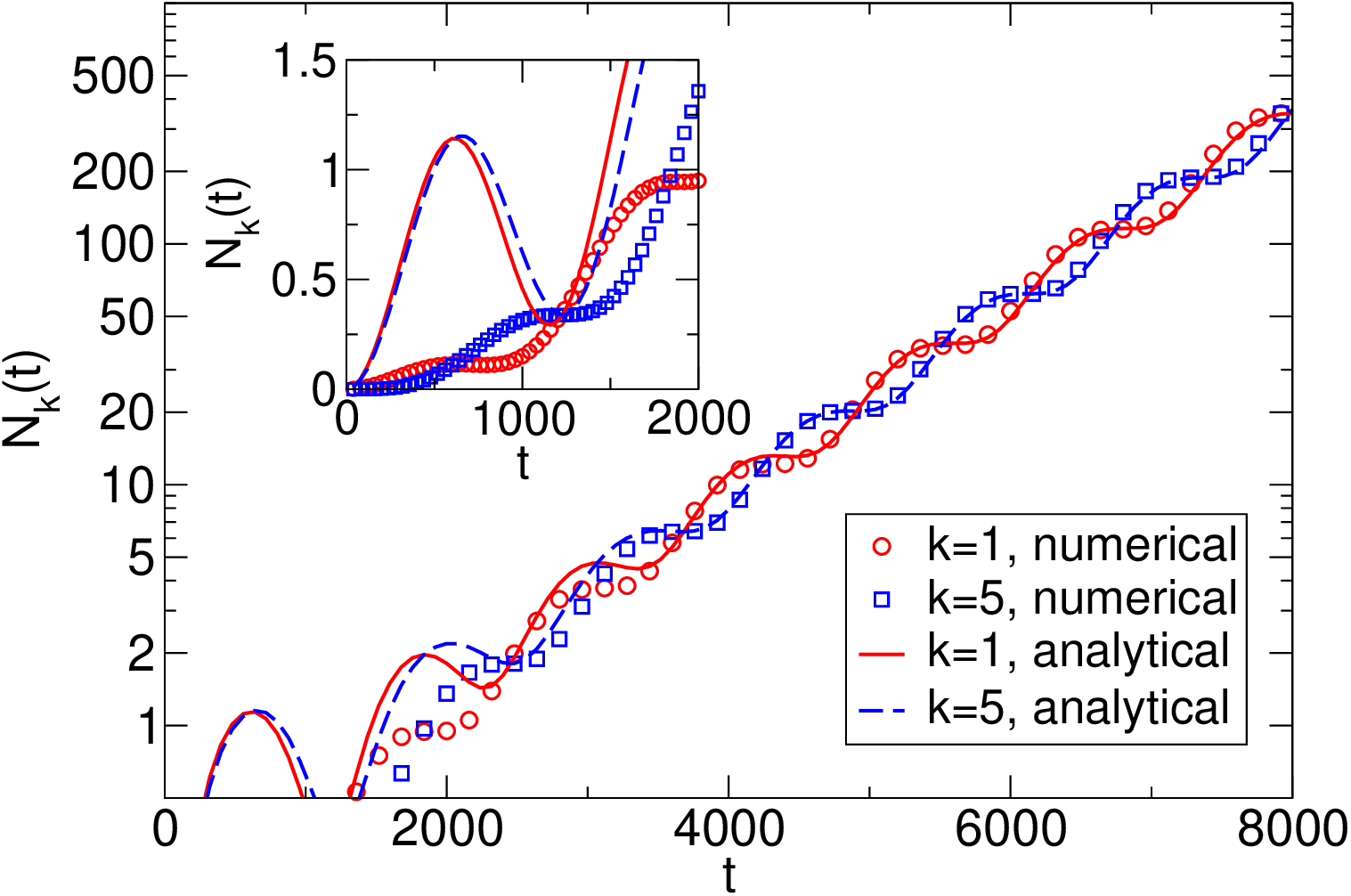}
\caption{(color online) Number of particles created in the modes $n=1$ and $k=5$ for 
$\omega=2\Omega_1^0$ and $M=\sqrt{2}\pi$ corresponding to the particle 
spectra depicted in Fig. \ref{figure 9}. The numerical results are compared to the 
analytical predictions Eq.~(54) [solid line] and Eq.~(55) [dashed line] of 
\cite{Crocce:2001}. The numerical results shown correspond to the cut-off parameter 
$k_{\rm max}=20$ which guarantees stability. 
\label{figure 10}}
\end{figure}

We now turn to cases with exact coupling between two modes. 
As already mentioned above exact coupling of modes takes place 
if the conditional equation (\ref{coupling condition}) has integer 
solutions. In particular, exact coupling between two modes $n$ and $k$
occurs if Eq.~(\ref{nk coupling}) is satisfied. In \cite{Crocce:2001} 
the authors derived analytical expressions (Eqs.~(54) and (55) of 
\cite{Crocce:2001}) for the case that the TE-mode $\Omega^0_{(1,1,1)}$ 
(resonant mode) is coupled to the mode $\Omega^0_{(5,1,1)}$, i.e. 
$3\Omega_{(1,1,1)}^0=\Omega_{(5,1,1)}^0$ is fulfilled. This particular
case is equivalent to the coupling of the massive modes $n=1$ and 
$k=5$ if $M=\sqrt{2}\pi$ $(l_0=1)$. Figure~\ref{figure 9} shows the 
numerically obtained particle spectrum at four different times. 
The cut-off parameter $k_{\rm max}=20$ guarantees stability of the 
numerical results.  
The numerical simulations confirm the prediction that practically only the modes 
$n=1$ and $k=5$ become excited and particles are produced  exclusively
in the two coupled modes. Thereby the rate of particle creation is
equal for the two modes. In Fig.~\ref{figure 10} we show the numerical 
results for $N_1(t)$ and $N_5(t)$ and compare them with the analytical 
expressions Eq.~(54) and Eq.~(55) of \cite{Crocce:2001} derived via
multiple scale analysis (MSA). 
Whereas the numerical results agree quite well with the analytical prediction of 
\cite{Crocce:2001} for long times, one observes a discrepancy between
the numerical results and the analytical predictions for "shorter times" up to 
$t\sim 3000$ ($\epsilon\pi\,t=3\pi$). For long times, the analytical predictions 
nicely reproduce the large scale oscillations in the exponentially
increasing particle numbers. For times up to $t\sim 500$, the
numerically calculated particle numbers grow with a much smaller 
rate than predicted by Eqs.~(54) and (55) of \cite{Crocce:2001}. Furthermore, the 
analytical expressions predict that for ``short times'' $N_1$ and 
$N_5$ increase with the same rate whereas from the numerical simulations we find 
that the production of particles in the mode $k=5$ sets in after the production of 
particles in the $n=1$-mode. Apart from the differences for short 
times the numerical results are well described by the analytical predictions of 
\cite{Crocce:2001}. The discrepancy between the analytical predictions
and the numerical results for short times is due to the fact that the
MSA analysis in \cite{Crocce:2001} only considers the resonant 
coupled modes, but for short enough times all modes should be treated on an equal  
footing \cite{Dalvit:private}.

\begin{figure}
\includegraphics[height=5.6cm]{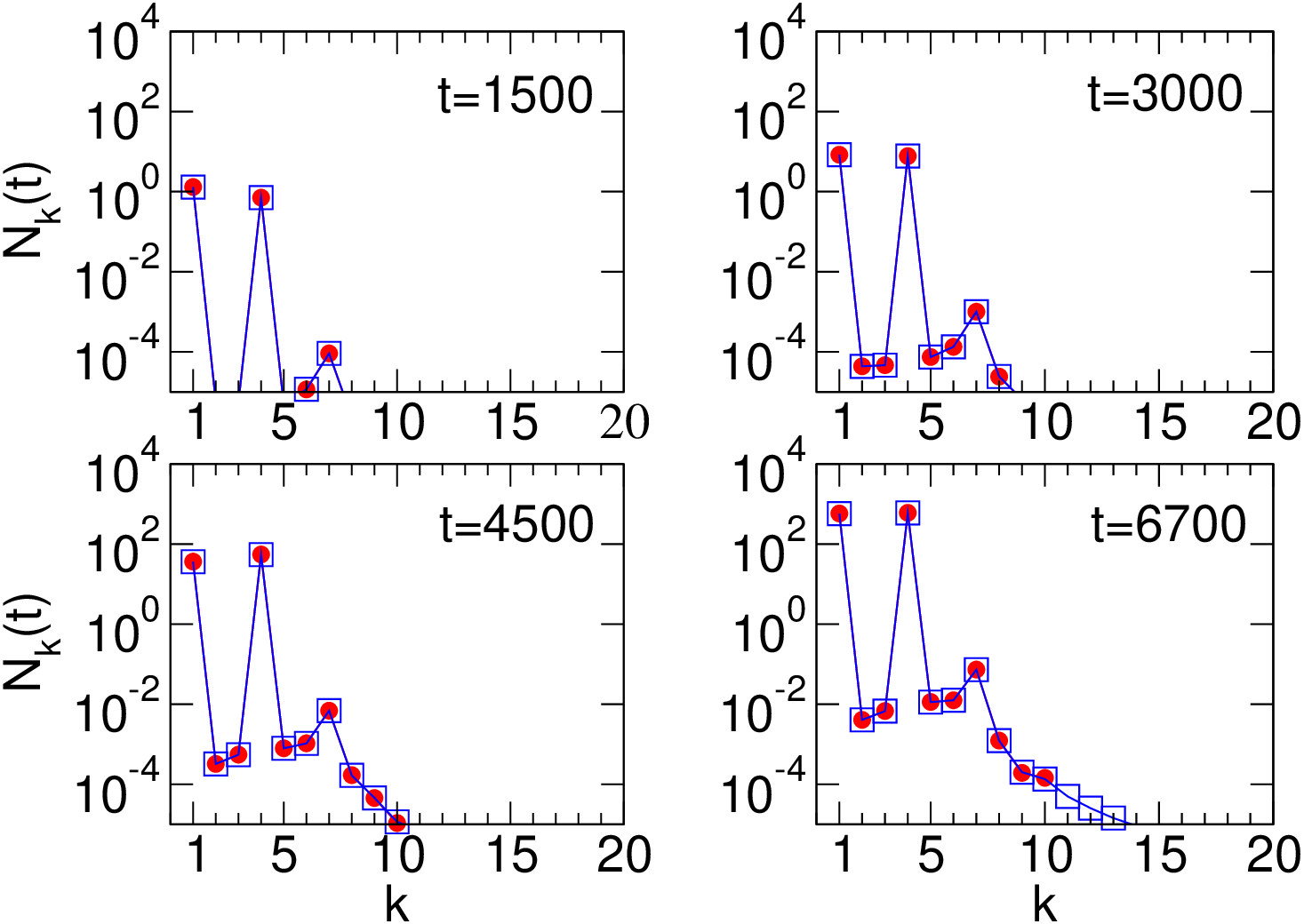}
\caption{(color online) Particle spectra for $\omega=2\Omega_1^0$ and 
mass parameter $M=\sqrt{7/8}\pi$ yielding exact coupling
between the modes $n=1$ and $k=4$. Dots correspond to 
$k_{\rm max}=10$ and squares to $k_{\rm max}=20$.
\label{figure 11}}
\end{figure}

\begin{figure}
\includegraphics[height=5.6cm]{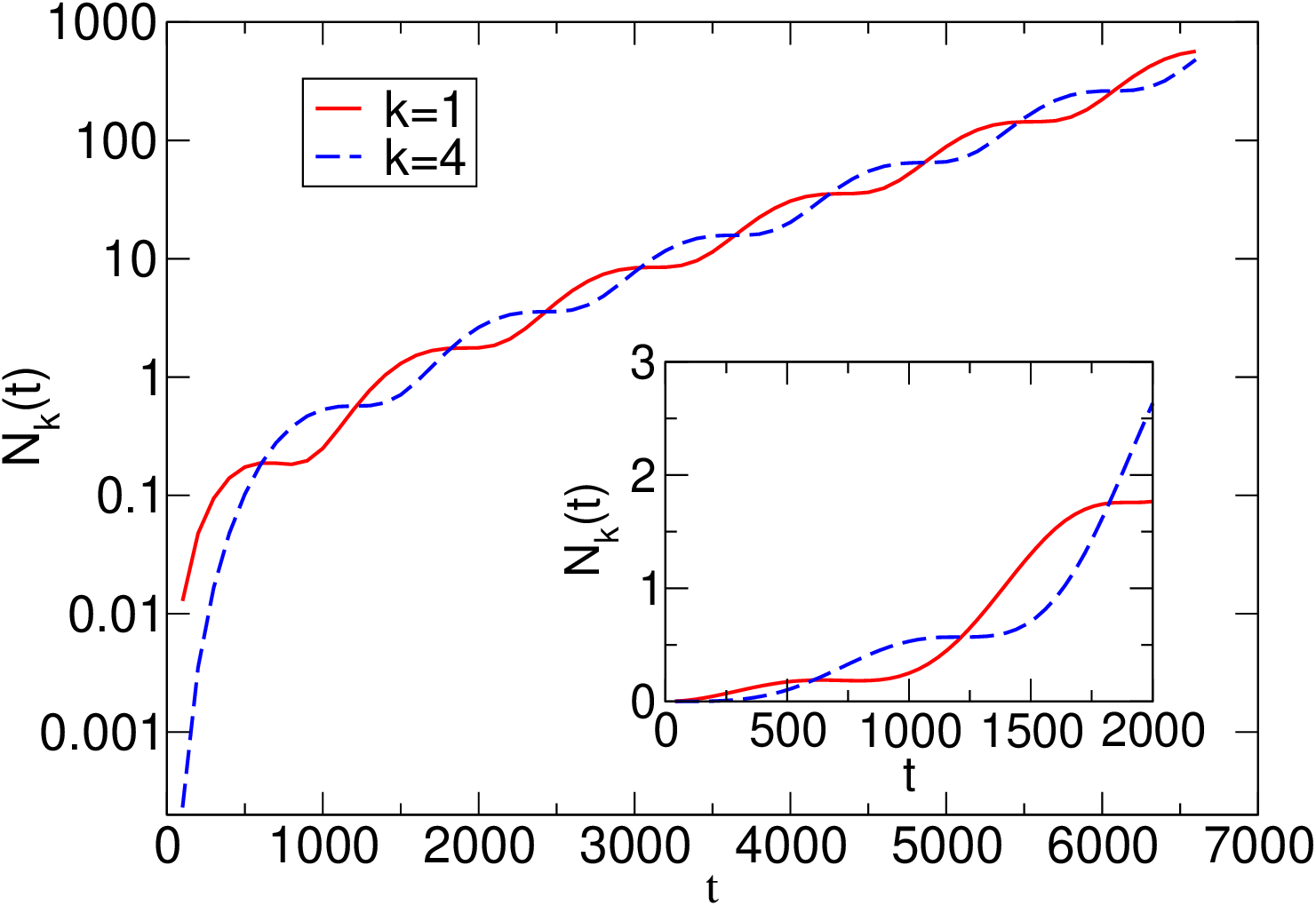}
\caption{(color online) Number of particles created in the modes $n=1$ and $k=4$ for 
$\omega=2\Omega_1^0$ and mass parameter $M=\sqrt{7/8}\pi$ 
corresponding to the particle spectra depicted in Fig. \ref{figure 11}.
\label{figure 12}}
\end{figure}

As a second example of exact coupling between two modes we show in 
Figs.~\ref{figure 11} and \ref{figure 12} the numerical results
obtained for $M=\sqrt{7/8}\pi$ for which the mode $n=1$ is 
coupled to the mode $k=4$. 

\begin{figure}
\includegraphics[height=5.6cm]{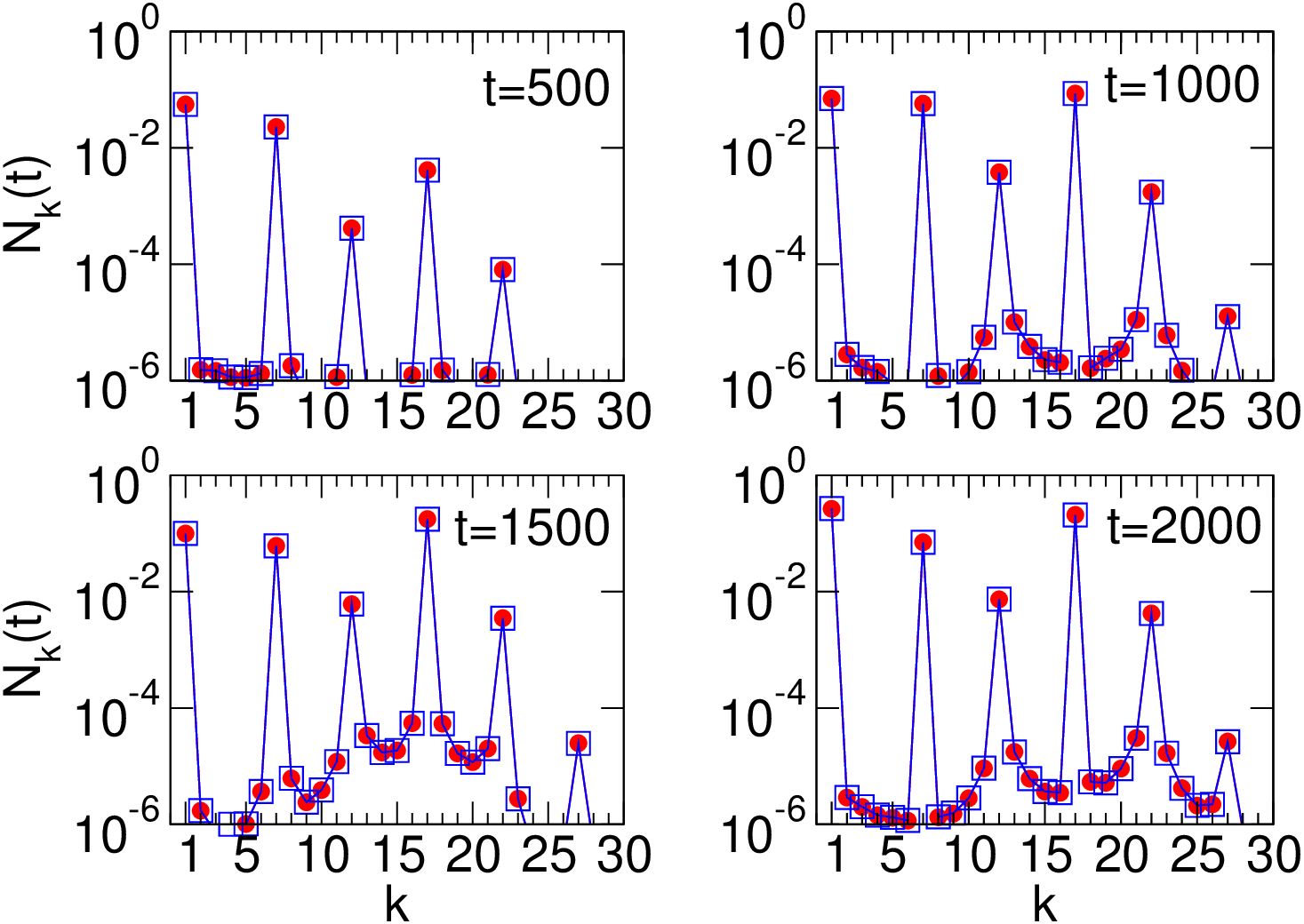}
\caption{(color online) Particle spectra for $\omega=2\Omega_1^0$ and mass parameter 
$M=\sqrt{5}\pi$. Dots correspond to $k_{\rm max}=40$ and squares to $k_{\rm max}=50$. 
\label{figure 13}}
\end{figure}

\begin{figure}
\includegraphics[height=5.6cm]{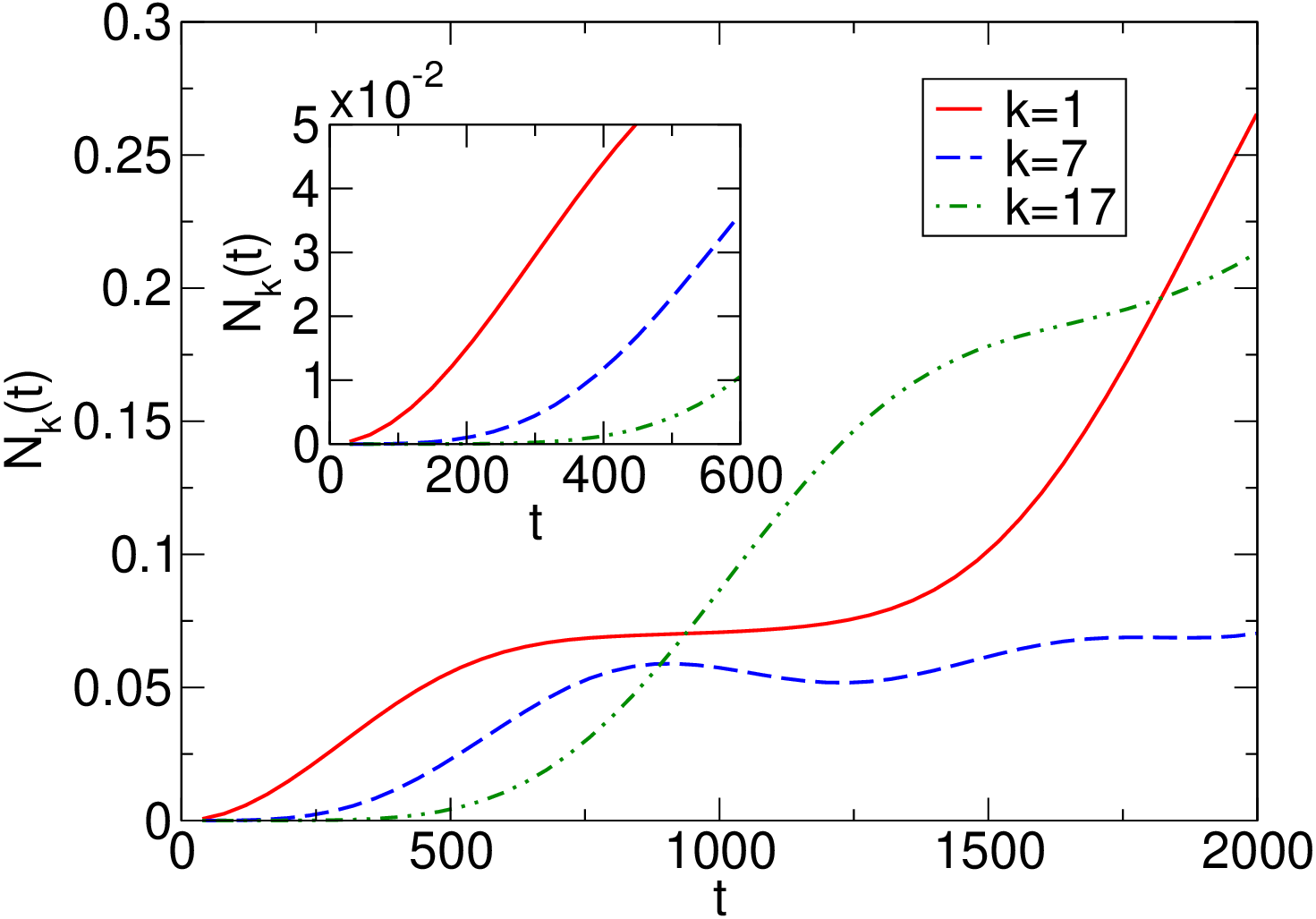}
\caption{(color online) Number of particles created in the 
modes $n=1$, $k=7$ and $l=17$ for $\omega=2\Omega_1^0$ 
and mass parameter $M=\sqrt{5}\pi$ corresponding to the 
particle spectra depicted in Fig.~\ref{figure 13}.
\label{figure 14}}
\end{figure}

Let us discuss another case with exact coupling of two modes which impressively 
demonstrates that strong coupling between modes $k$ and $l$ occurs even if 
Eq.~(\ref{coupling condition}) is satisfied only approximately. 
For $M=\sqrt{5}\pi$ equation (\ref{coupling condition}) predicts that the 
mode $n=1$ is exactly coupled to the mode $k=7$ (i.e. $k=7$ is an integer solution of 
$3\Omega_1^0=\Omega_k^0$). The equation
$2\Omega_1^0=\Omega_l^0-\Omega_7^0$ is not satisfied by an integer $l$
but has the solution $l\sim 12.04$ which is close to the integer
$l=12$. Thus we can expect a coupling of the mode $k=7$ to the 
mode $l=12$. In addition one finds that the equation 
$2\Omega_1^0 = \Omega_m^0-\Omega_{12}^0$ has solution $m=16.96$, i.e. $m\sim 17$,  
and hence $l=12$ is coupled to $m=17$. In the same way the equation 
$2\Omega_1^0 = \Omega_j^0-\Omega_{17}^0$ which is solved by $j \sim 21.93$
leads to a coupling between the modes $m=17$ and $j=22$. 
Hence from the numerical simulations we expect to find a particle
spectrum showing that particle creation takes place in the 
modes $k=1,7,12,17$ and $22$. This is demonstrated in 
Fig.~\ref{figure 13} where the numerically evaluated particle spectrum is depicted 
for times $t=500, 1000, 1500$ and $2000$. The cut-off parameter $k_{\rm max}=50$ 
ensures numerical stability in the integration range considered
\footnote{ 
From Fig.~\ref{figure 13} one observes that also the mode $l=27$
is weakly coupled. The equation $2\Omega_1^0 =
\Omega_l^0-\Omega_{22}^0$  has the solution $l\sim 26.92$ 
which explains the excitation of the mode $l=27$.
}. The number of created particles $N_k(t)$ is shown in Fig.~\ref{figure 14}
for the modes $k=1,7$ and $17$. 

Without having done a detailed analysis we find, as a reasonable 
approximation, that a mode $l$ is (strongly) coupled to a given 
mode $k$ whenever the ratio 
$|l-\tilde{l}|/l$ with $\tilde{l}$ denoting the solution of 
$2\Omega_n^0=|\Omega_{\tilde{l}}^0 \pm \Omega_k^0|$
is of the order of or smaller than $10^{-3}$, i.e. of the order of or
smaller than $\epsilon$ used in the simulations.
 
\subsection{Higher resonance $\omega=2\Omega_2^0$}

\begin{figure}
\includegraphics[height=5.6cm]{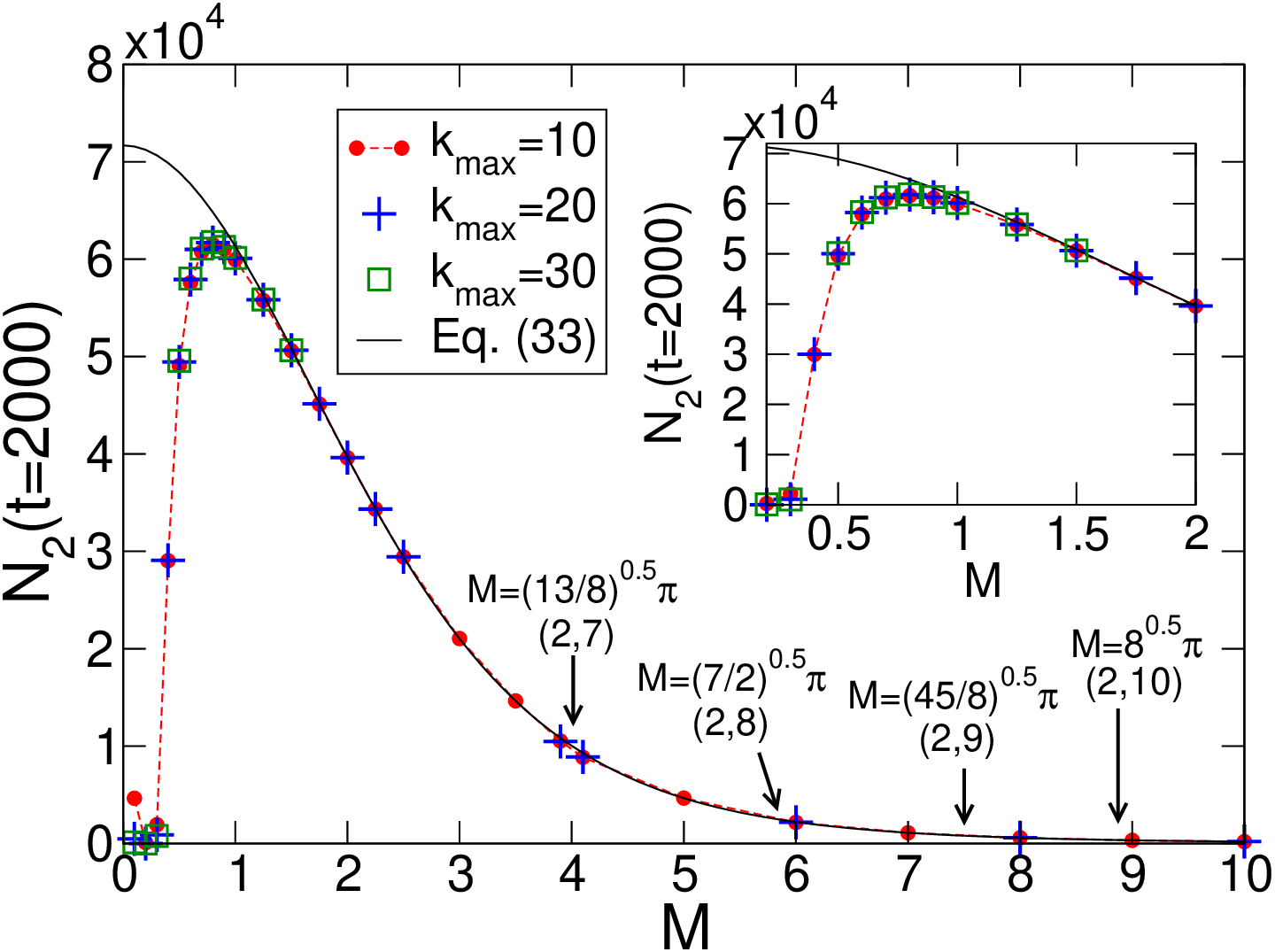}
\caption{(color online) Number of particles created in the resonance mode $n=2$ 
as function of the mass parameter $M$. The solid line corresponds to the 
analytical prediction (\ref{massive particle number}). Note that the first three 
values in the spectrum are not numerically stable due to an insufficient $k_{\rm max}$. 
\label{figure 15}}
\end{figure}

Now we briefly discuss results obtained for the cavity frequency 
$\omega=2\Omega_2^0$. In Fig.~\ref{figure 15} we show a numerically 
calculated mass spectrum similar to the one depicted in Fig.~\ref{figure 5}. 
The qualitative behavior is the same as discussed for the 
main resonance case. As in Fig.~\ref{figure 5} results for values of
the mass parameter $M$ for which modes are exactly coupled are not
included in the spectrum but marked by arrows with 
the corresponding coupled modes given in brackets. The numerical results again 
perfectly agree with the analytical prediction (\ref{massive particle number}) for 
values of $M$ larger than a threshold value which is roughly $1.3$. The maximum in 
the particle spectrum appears now for $M\sim 0.8$ and the interpretation of 
the shape of the mass spectrum is equivalent to the one given for the case 
$\omega=2\Omega_1^0$.    

\begin{figure}
\includegraphics[height=5.6cm]{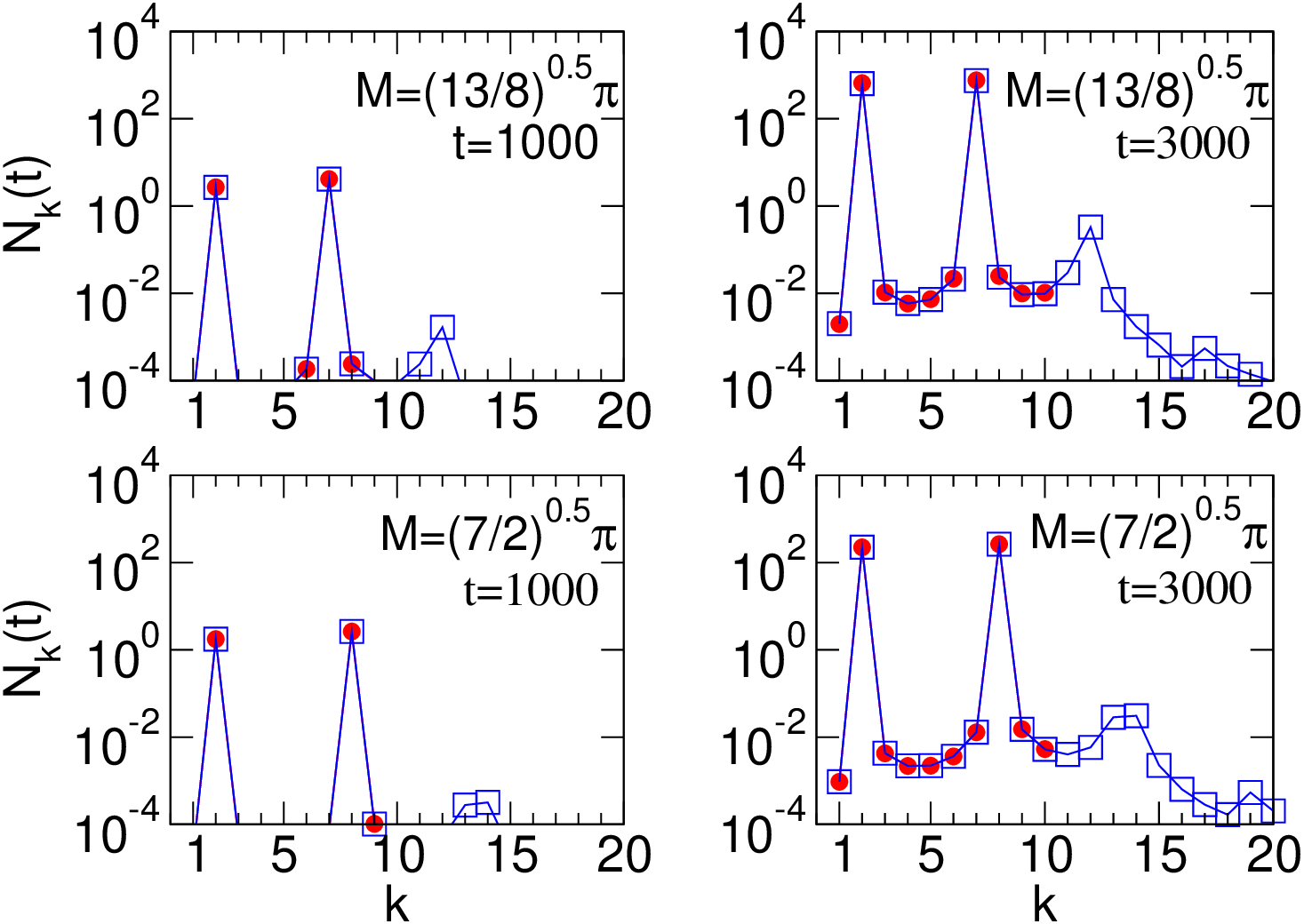}
\caption{(color online) Particle spectra for $\omega=2\Omega_2^0$ and mass parameters 
$M=\sqrt{13/8}\pi$ and $M=\sqrt{7/2}\pi$. 
\label{figure 16}}
\end{figure}

\begin{figure}
\includegraphics[height=5.6cm]{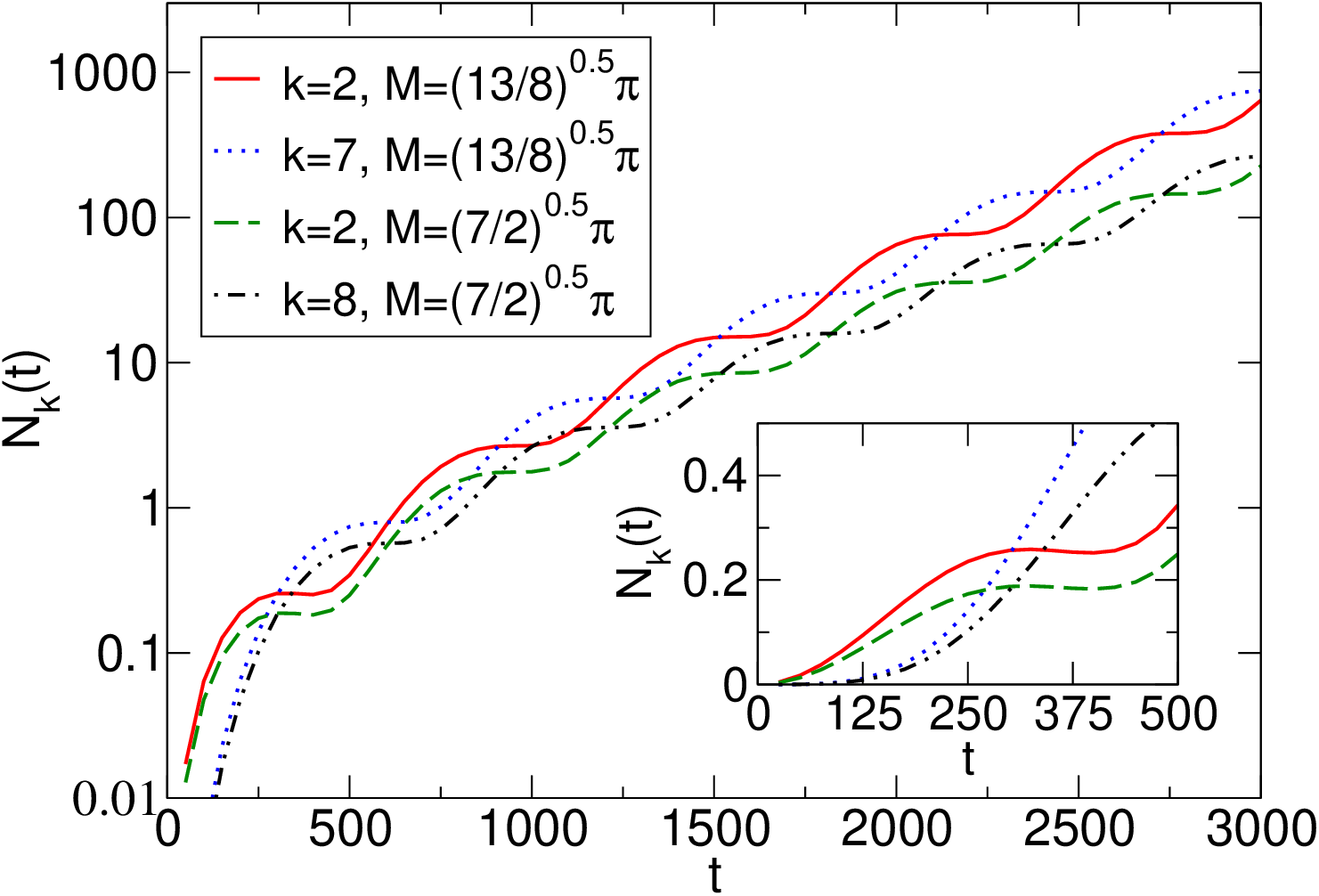}
\caption{(color online) Number of particles created in the modes $k=2$ and $7$ for 
$M=\sqrt{13/8}\pi$, respectively, $k=2$ and $8$ for 
$M=\sqrt{7/2}\pi$, corresponding to the particle spectra shown in 
Fig.~\ref{figure 16}. 
\label{figure 17}}
\end{figure}

In Figs.~\ref{figure 16} and \ref{figure 17} we finally show 
numerical results for the two mass parameters $M=\sqrt{13/8}\pi$
and $M=\sqrt{7/2}\pi$ yielding exact coupling between the modes $2$
and $7$, respectively, $2$ and $8$.

\section{Photon creation in a three-dimensional cavity}
The analogy between massive scalar particles and transverse
electric photons in a three-dimensional rectangular cavity 
outlined in Section II allows to interpret the presented numerical 
results as follows: Consider a three-dimensional rectangular cavity 
with equally sized non-dynamical dimensions $l_y=l_z \equiv l_\|$. 
We parametrize the size of $l_\|$ in terms of the initial size of 
the dynamical dimension $l_0=l_x(0)$ by introducing
$\ell=l_\|/l_0$. If we restrict ourselves for simplicity
to the case $n_y=n_z\equiv n_\|$, the dimensionless mass parameter 
$M$ reads
\begin{equation}
M=l_0 k_\|=\sqrt{2}\left(\frac{n_\|\pi}{\ell}\right).
\label{mass cavity relation}
\end{equation}
Therefore, for fixed $n_\|$, any value of $M$ corresponds to a
particular realization, i.e. size $\ell$, of the non-dynamical cavity
dimensions. We have found that for a particular value $M$ the 
production of massive scalar particles in the resonant mode is
maximal. Consequently it is possible to maximize the production 
of TE-photons in a three-dimensional rectangular cavity by 
tuning the size $\ell$ of the non-dynamical cavity dimensions. 

For instance, for $\omega=2\Omega_1^0$ the creation of massive 
scalar particles in the resonant mode $n=1$ is most efficient 
for $M\sim 0.4$ (cf. Fig. \ref{figure 5}). This corresponds to 
the three-dimensional case with $\omega=2\Omega_{(1,1,1)}^0$ and 
$\ell\sim 11$. Hence by designing the three-dimensional cavity 
such that $l_\|\sim 11\,l_0$ the production of TE-mode photons 
in the resonant mode $(1,1,1)$ can be maximized.
In order to maximize the creation of TE-photons in the 
mode $(1,2,2)$ when $\omega=2\Omega^0_{(1,2,2)}$ 
the size $l_\|$ of the non-dynamical dimensions has to be 
doubled, i.e $l_\|\sim 22\, l_0$. For $\omega=2\Omega_2^0$ 
we have found that the maximum in the mass spectrum is at $M\sim 0.8$ 
[cf Fig.~15]. Accordingly, the production of TE-photons of frequency 
$\Omega^0_{(2,1,1)}$ under resonance conditions is maximal in a 
cavity of dimensions $l_\| \sim 5.6 \,l_0$. 
The strong-coupling case $\omega=2\Omega_1^0$ with $M=0.2$ 
where the analytical prediction (\ref{massive particle number}) does not 
describe the numerical results due to enhanced intermode coupling 
[cf Figs. \ref{figure 1} and \ref{figure 2} ] 
corresponds to the lowest TE-mode $(1,1,1)$ in a cavity 
of size $l_\|\sim 22\,l_0$. 

Similarly one can arrange the size of the cavity such that particular
modes are exactly coupled, i.e. Eq.~(\ref{nk coupling}) is satisfied. 
For instance, resonant coupling of the TE-modes $(1,1,1)$ and
$(4,1,1)$ corresponds to $\omega=2\Omega_1^0$ with $M=\sqrt{7/8}\pi$ 
[cf Figs.~11 and~12] and is therefore realized in a cavity of size  
$l_\|\sim 1.5\,l_0$. Finally, choosing $l_\|\sim 0.63 \,l_0$, i.e. 
$M=\sqrt{5}\pi$, couples the TE-modes $(1,1,1)$, $(7,1,1)$,$(12,1,1)$, 
$(17,1,1)$ and $(22,1,1)$ in the resonance case 
$\omega=2\Omega^0_{(1,1,1)}$ [cf Figs.13 and~14]. 

In summary, the mass spectrum Fig. \ref{figure 5} can be interpreted in the
following way if we set $n_\|=1$ and $\omega=2\Omega^0_{(1,1,1)}$: 
For $l_\|=l_0$, i.e. cubic cavity, 
the modes $(1,1,1)$ and $(5,1,1)$ are resonantly coupled 
(cf. Figs. \ref{figure 9} and \ref{figure 10}). Enlarging $l_\|$ with respect to $l_0$
increases the production of resonance mode photons $(1,1,1)$ 
until $l_\|=1.5\,l_0$ ($M=\sqrt{7/8}\pi$) is approached where the modes 
$(1,1,1)$ and $(4,1,1)$ are exactly coupled (cf. Figs. \ref{figure 11}
and \ref{figure 12}). When increasing $l_\|$ further photon creation
in the TE-mode $(1,1,1)$ becomes more and more efficient and is
perfectly described by Eq.~(\ref{massive particle number}). 
Reaching $l_\|\sim 7.4 \,l_0$ ($M\sim0.6$, the threshold) the intermode 
coupling starts to become noticeable causing slight deviations of the
numerical results from the analytical prediction. For 
$l_\|\sim 11 \,l_0$ $(M\sim 0.4)$ the production of TE-mode photons
is most efficient. The number of Photons created in the mode
$(1,1,1)$ is smaller than the analytical prediction 
Eq.~(\ref{massive particle number}) because of the 
coupling of the modes $(1,1,1)$ and $(3,1,1)$. When increasing $l_\|$ 
beyond $\sim 11 \,l_0$ the strength of the intermode coupling 
is enhanced drastically and consequently the number of 
produced TE-mode photons decreases rapidly. For  $l_\|\sim 22 \,l_0$, for
instance, the mode $(1,1,1)$ is (strongly) coupled to the modes 
$(3,1,1)$ and $(5,1,1)$ (cf Fig. \ref{figure 2}).
Reducing $l_\|$ with respect to $l_0$ (i.e. going
to masses $M>\sqrt{2}\pi$) lowers the efficiency of photon creation in
the resonant mode. The mass spectrum Fig. \ref{figure 15} owns
an equivalent interpretation.   

\section{Conclusions}
The production of massive scalar particles in a one-dimensional 
cavity, or analogously the creation of TE-mode photons in a 
three-dimensional rectangular cavity, has been studied numerically 
for resonant wall oscillations. 

We have found perfect agreement between the numerical 
results and analytical predictions of \cite{Crocce:2001} in the 
case that no modes are (strongly) coupled. When two modes are exactly 
coupled, i.e. Eq.~(\ref{coupling condition}) possesses integer
solutions for $l$ and $k$, the numerical results agree with analytical 
predictions of \cite{Crocce:2001} for sufficiently long times
but disagree for short times. The discrepancy for short times 
is ascribed to properties of the multiple scale analysis 
used in \cite{Crocce:2001}. 

The effect of the intermode coupling has been studied in detail which
is only possible by means of numerical simulations. 
As main result we have found that a particular mass exists for which the
production of massive scalar particles is most efficient. The
appearance of a maximum in the mass spectrum, i.e. the number of
created particles after a given time as function of mass, 
is explained by the increasing strength of the intermode coupling when 
decreasing the mass below a certain threshold value. 

The analogy between massive scalar particles in a one-dimensional cavity and
TE-mode photons in a three-dimensional cavity allows the conclusion
that the efficiency of TE-mode photon production from vacuum  
in a resonantly vibrating rectangular cavity can be controlled 
(and maximized) by tuning the size of the cavity when keeping the 
quantum numbers $n_y,n_z$ corresponding to the non-dynamical cavity
dimensions fixed. 

The main resonance case $\omega=2\Omega^0_{(1,1,1)}$ has been
discussed for a cavity with equally sized non-dynamical dimensions
$l_\|=l_y=l_z$ in detail in section VI. We have shown that photon 
creation in the resonant mode $(1,1,1)$ is most
efficient if the size $l_\|$ of the non-dynamical cavity dimensions 
is $\sim 11$ times larger than the dynamical cavity dimension. 
The existence of a certain cavity size which maximizes photon creation
is among other things explained by the fact that intermode 
coupling takes place even if Eq. (\ref{coupling condition}) 
is satisfied only approximately. If $l_\|$ is larger than 
this value the intermode coupling is so strong that higher frequency
modes like $(3,1,1)$ and $(5,1,1)$ couple to the resonant mode
$(1,1,1)$ and strongly damp its evolution.
Furthermore the coupling of particular field modes by tuning the size
of the cavity has been studied. The effects provoked by the
intermode coupling can be studied in full detail only by means of numerical methods.  

Our findings demonstrate that the intermode coupling in dynamical cavities
plays an important role. Even if analytical results are known, numerical 
simulations are a very useful and indeed necessary tool because only they  
can completely take into account the intermode coupling. 
In order to study photon creation associated with the full electromagnetic 
field in a dynamical cavity also the contribution of the transverse 
magnetic modes (TM) to the photon production has to be considered. 
Studying TM-modes numerically represents a more
demanding task because of the more complicated so-called generalized
Neumann boundary condition these modes are subject to. This will be
addressed in a future work.

\begin{acknowledgments}
The author is grateful to Ruth Durrer and Cyril Cartier for valuable discussions, 
carefully reading of the manuscript and useful comments. He would also like to 
thank Ralf Sch{\"u}tzhold and G{\"u}nter Plunien for discussions and
comments on the manuscript. Furthermore the 
author is much obliged to Diego Dalvit and Emil Mottola for enlightening and 
interesting discussions as well as their kind hospitality during his visit to the 
Los Alamos National Laboratory. Finally, the author would like to thank 
Paulo Maia Neto and Francisco Mazzitelli for discussions and comments during 
the Seventh Workshop On Quantum Field Theory Under The Influence Of External 
Conditions, Barcelona, Spain, 2005. Financial support from the Swiss National 
Science Foundation is gratefully acknowledged.

\end{acknowledgments}

\begin{appendix}
\section{Remarks on numerics}
To solve the system of differential equations formed by Eqs.~(\ref{deq for xi}) and 
(\ref{deq for eta}) numerically we decompose $\xi_n^{(m)}(t)$ and $\eta_n^{(m)}(t)$ 
in their real and imaginary parts:
\begin{equation}
\xi_n^{(m)}=u_n^{(m)} + i v_n^{(m)}\,,\,\,\eta_n^{(m)}=x_n^{(m)} + i y_n^{(m)} .
\label{decomp of xi and eta}
\end{equation}
The resulting coupled system of first-order differential equations can then be 
written in the form
\begin{equation}
\underline{\dot{X}}^{(m)}(t)=\underline{W}(t)\underline{X}^{(m)}(t)
\label{num deq system}
\end{equation}
with real vectors $\underline{X}^{(m)}(t)$ and matrix $\underline{W}(t)$.
Choosing the representation 
\begin{equation}
\underline{X}^{(m)}=(u_1^{(m)}..u_{K}^{(m)}x_1^{(m)}..x_{K}^{(m)}
v_1^{(m)}..v_{K}^{(m)}y_1^{(m)}..y_{K}^{(m)})^T,
\end{equation}
where we have truncated the infinite system via introducing the cut-off 
parameter $K\equiv k_{\rm max}$, the  $4K \times 4K$ - matrix 
$\underline{W}(t)$ becomes
\begin{equation}
\underbar{W}(t)=-
\left[\begin{array}{cccc}
C^-(t)&C^+(t)&-A^+(t)&A^-(t)\\
C^+(t)&C^-(t)&-A^-(t)&A^+(t)\\
A^+(t)&-A^-(t)&C^-(t)&C^+(t)\\
A^-(t)&-A^+(t)&C^+(t)&C^-(t)
\end{array}\right]
\end{equation}
with the $K\times K$ - matrices
$C^{\pm }(t)=\left[c_{kn}^{\pm}(t)\right]$, $1\leq k,n \le K$ and
diagonal matrices $A^{\pm}(t)=\left[ a_{nn}^\pm(t)\right]$
where $a_{nn}^\pm(t)$ and $c_{nk}^\pm(t)$ are defined in Eq.~(\ref{def a matrix}) 
and Eq.~(\ref{def c matrix}), respectively. The number of particles 
(\ref{particle number}) created in a mode $n$ 
at $t=t_1$ may now be expressed in terms of the real functions: 
\begin{eqnarray}
N_n(t_1)&=&\frac{1}{4}\sum_{m=1}^{K}\frac{\Omega_n^1}{\Omega_m^0} \Big 
\{ \left[ \Delta_n^-(t_1)u_n^{(m)}(t_1)+
\Delta_n^+(t_1)x_n^{(m)}(t_1)\right]^2 \nonumber\\
&+&\left[ \Delta_n^-(t_1)v_n^{(m)}(t_1)+\Delta_n^+(t_1)y_n^{(m)}(t_1)\right]^2
\Big \}
\label{num particle number}
\end{eqnarray}
which in the particular case $t_1={\cal N}\,T$ with $T$ the period of the
cavity oscillations and integer ${\cal N}$ reduces to 
\begin{equation}
N_n({\cal N}\,T)=\frac{1}{4}\sum_{m=1}^K\left[\left(x_n^{(m)}({\cal N}\,T)\right)^2+
\left(y_n^{(m)}({\cal N}\,T)\right)^2\right].
\label{num particle number two}
\end{equation} 
In order to calculate (\ref{num particle number}) the system (\ref{num deq system}) 
has to be evolved numerically $K$-times ($m$ is running from $1$ to $K=k_{\rm max}$) 
up to $t=t_1$ with initial conditions
\begin{equation}
v_n^{(m)}(0)=x_n^{(m)}(0)=y_n^{(m)}(0) =0
\end{equation}
and
\begin{equation}
u_n^{(m)}(0)=2\delta_{nm}.
\end{equation}
Besides investigating the stability of the numerical solutions in 
dependence on the cut-off $K$ the quality of the numerical solutions 
can be assessed by checking the validity of the Bogoliubov relations 
\begin{equation}
\sum_m\left[{\cal A}_{mn}(t_1) {\cal A}^*_{mk}(t_1)-{\cal B}^*_{mn}(t_1)
{\cal B}_{mk}(t_1)\right]=\delta_{nk}
\label{Bogoliubov 1}
\end{equation}
\begin{equation}
\sum_m\left[{\cal A}_{mn}(t_1) {\cal B}^*_{mk}(t_1)-{\cal B}^*_{mn}(t_1)
{\cal A}_{mk}(t_1)\right]=0.
\end{equation}
In order to solve the system (\ref{num deq system}) numerically we applied 
integration routines based on different standard solvers. Mainly employed were 
the Runge-Kutta-Fehlberg 4th-5th order method ({\tt rkf45}) and the Runge-Kutta  
Prince-Dormand method ({\tt rk8pd}). Source codes provided by the GNU Scientific 
Library (GSL) \cite{gsl} as well as the MATPACK - Library \cite{matpack} 
were used. 

\begin{figure}
\includegraphics[height=6.2cm]{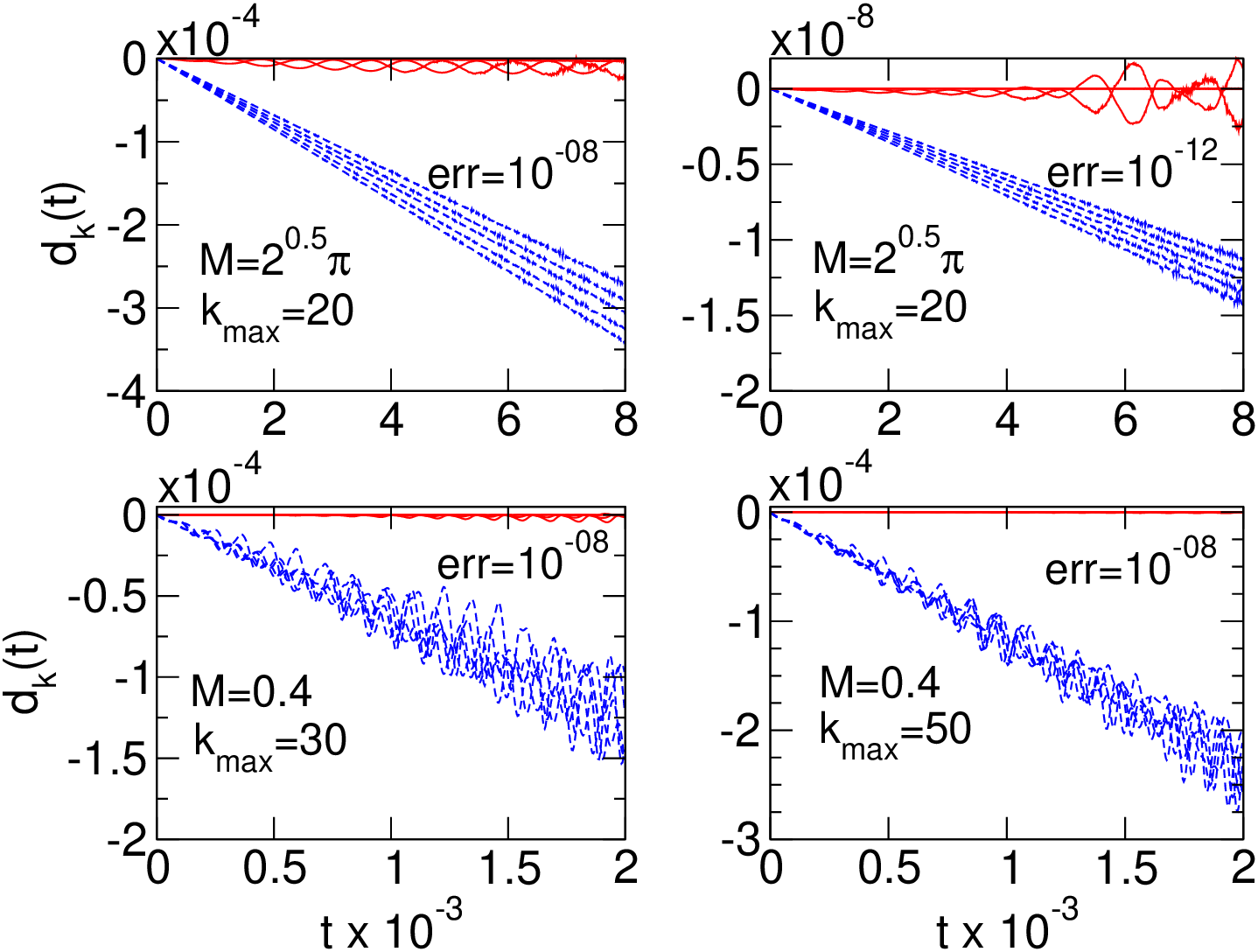}
\caption{(color online) The function $d_k(t)$ [Eq. (\ref{d function})] 
for $\omega=2\Omega_1^0$ and $M=\sqrt{2}\pi$ [panel (a) and (b)] and
$M=0.4$ [panel (c) and (d)]. In any case $d_k(t)$ is shown for
$k=1,...,5$ (upper bands) and the last five values 
$k=k_{\rm max}-4,...,k_{\rm max}$ (lower bands). With ``err'' we
denote the presetted values for the relative and absolute error
used in the numerical simulations performed with, in these cases, the
Runge-Kutta Prince-Dormand method.
\label{figure 18}}
\end{figure}

In the following we discuss the accuracy of the numerical simulations 
by considering the quantity
\begin{equation}
d_k(t)=1-\sum\left(|{\cal A}_{mk}(t)|^2 - |{\cal B}_{mk}(t)|^2\right),
\label{d function}
\end{equation}
indicating to what accuracy the diagonal part of the relation 
(\ref{Bogoliubov 1}) is satisfied by the numerical solutions. 
Generic examples for $d_k(t)$ are shown in Fig.~\ref{figure 18}. 
Panels~(a) and (b) correspond to the exact coupling case
$M=\sqrt{2}\pi$ (cf. Figs. \ref{figure 9} and \ref{figure 10})
with the absolute and relative errors (err) 
for the Runge-Kutta Prince-Dormand method ({\tt rk8pd}) \cite{gsl}  
have been set to $10^{-8}$ (a) and $10^{-12}$ (b).   
Thereby two ``bands'' are shown. The upper one correspond to $k=1$ to
$5$ whereas the lower one correspond to $k=16$ to $k_{\rm max}=20$.
The deviation from zero is larger for higher $k$ because these modes
are more affected by the truncation of the infinite system through the cut-off
$k_{\rm max}$. Comparing the absolute value of the maximal deviation
of $d_k(t=8000)$ from zero which is $\sim 3.5\times 10^{-4}$ 
for err=$10^{-8}$ and $\sim 1.5\times 10^{-8}$ for err=$10^{-12}$
with the number of particles created in the resonantly excited modes 
$N_1(t=8000)\sim 350$ and $N_5(t=8000)\sim 400$ demonstrates that  
the numerical simulations guarantee a good accuracy.     

In panel (c) and (d) of Fig.~\ref{figure 18} we show $d_k(t)$ for the case $M=0.4$
(cf. Figs. \ref{figure 6} and \ref{figure 7}) for the cut-off 
values $k_{\rm max}=30$ (c) and $k_{\rm max}=50$ (d). The numerical
simulations have been performed with err=$10^{-8}$ and again two bands
are shown corresponding to the first five (upper band) and last five 
(lower band) values of $k$. For $k_{\rm max}=50$ the deviation of the 
absolute value of $d_k$ from zero for the last values $k=46,...,50$
[panel (d)] is slightly larger compared to the deviation for
the last five modes for $k_{\rm max}=30$. But the deviation of
$d_k(t)$ from zero for the first modes $k=1,...,5$ is smaller for 
$k_{\rm max}=50$ than for $k_{\rm max}=30$, i.e. the accuracy for the 
first modes improves when increasing $k_{\rm max}$ as it is expected. 
Comparing $|d_k(t=2000)|\sim 3\times 10^{-4}$ for $k_{\rm max}=50$ 
with the number of created particles $N_1(t=2000)\sim 124$ 
demonstrates again the accuracy of the numerical
simulations. The remaining Bogoliubov relations are satisfied with the
same accuracy.

\end{appendix}



\end{document}